\documentclass[preprint2]{aastex6}

\usepackage{amsmath}

\shortauthors{Sheehan et al.}
\shorttitle{Taurus Class I Disk Masses}

\begin{document}

\title{Disk Masses for Embedded Class I Protostars in the Taurus Molecular Cloud}
\author{Patrick D. Sheehan\altaffilmark{1} and Josh A. Eisner\altaffilmark{1}}
\affil{\altaffilmark{1}Steward Observatory, University of Arizona, 933 N. Cherry Avenue, Tucson, AZ, 85721}

\begin{abstract}
Class I protostars are thought to represent an early stage in the lifetime of protoplanetary disks, when they are still embedded in their natal envelope. Here we measure the disk masses of 10 Class I protostars in the Taurus Molecular Cloud to constrain the initial mass budget for forming planets in disks. We use radiative transfer modeling to produce synthetic protostar observations and fit the models to a multi-wavelength dataset using a Markov Chain Monte Carlo fitting procedure. We fit these models  simultaneously to our new CARMA 1.3 mm observations that are sensitive to the wide range of spatial scales that are expected from protostellar disks and envelopes so as to be able to distinguish each component, as well as broadband spectral energy distributions compiled from the literature. We find a median disk mass of 0.018 M$_{\odot}$ on average, more massive than the Taurus Class II disks, which have median disk mass of $\sim0.0025$ M$_{\odot}$. This decrease in disk mass can be explained if dust grains have grown by a factor of 75 in grain size, indicating that by the Class II stage, at a few Myr, a significant amount of dust grain processing has occurred. However, there is evidence that significant dust processing has occurred even during the Class I stage, so it is likely that the initial mass budget is higher than the value quoted here.
\end{abstract}


\section{Introduction}

Stars form from clouds of gas and dust that collapse under the strength of gravity. Conservation of angular momentum causes the majority of the material to be deposited into a circumstellar disk. Viscosity in the disk causes material to accrete onto the star. The viscous time in these disks is comparable to theoretical expectations of planet formation timescales.

Young stars have historically been classified by their near-infrared spectral index \citep{Lada1987,Myers1987,Andre1993} and bolometric temperature \citep[e.g.][]{Myers1993,Chen1995}. Class 0 protostars are characterized by a lack of optical and near-/mid-infrared emission, and low bolometric temperatures, suggesting that the central source is highly extincted. They are thought to represent the earliest stage of star formation, where a massive protostellar envelope shrouds the central protostar, obscuring it's light from view. They are likely forming disks as material from the envelope is funneled onto the protostar \citep{Ulrich1976,Terebey1984}. It is not clear whether these sources have rotationally supported disks, or whether magnetic braking at these early ages inhibits disk formation \citep[e.g.][]{Allen2003,Mellon2008,Li2013}. Rotationally supported disks have been observed around some Class 0 protostars \citep{Tobin2012,Tobin2013,Murillo2013,Codella2014,Lindberg2014,Aso2015}. 

Class I protostars are characterized by steeply rising near-infrared emission that peaks at mid-infrared wavelengths, and have bolometric temperatures of a few hundred Kelvin. They are likely sources with mature protoplanetary disks that are still being fed by a collapsing envelope of material \citep[e.g.][]{Harsono2014,Aso2015}. 

Class II YSO's have SEDs that are flat or declining at near-infrared wavelengths, with some light from the central star visible. By this stage the material in the envelope is thought to have been depleted onto the disk and protostar, exposing the stellar photosphere to observers. Finally Class III protostars are dominated by the light of the central protostar with a small amount of infrared excess, and are thought to be disks in which the gas has been depleted and only a small amount of rocky material remains.

Previous studies have shown this classification scheme to be prone to errors. For example it is possible to mistake an edge-on disk as a highly obscured Class I protostar \citep[e.g.][]{Chiang1999,Crapsi2008}. Disks that are highly obscured by foreground material have also been mistaken for Class I disks \citep[e.g.][]{Brown2012}. More recent studies have attempted to define other metrics for determining the evolutionary state of protostars, for example, based on bolometric temperatures and the strength of HCO$^{+}$ emission towards the source \citep[e.g.][]{vanKempen2009}. The best way, however, to probe the underlying density distribution is through spatially resolved observations of optically thin matter. Detailed radiative transfer modeling of datasets at multiple wavelengths can be used break model degeneracies, constrain parameters like temperature and opacity, and determine physical properties of the system \citep[e.g.][]{Osorio2003,Wolf2003,Eisner2005,Lommen2008,Gramajo2010,Eisner2012,Sheehan2014,Sheehan2017a}.

\floattable
\begin{deluxetable}{lccc}
\scriptsize
\tablenum{1}
\tablecaption{Log of CARMA Observations}
\label{table:obs}
\tablehead{\colhead{Source} & \colhead{Observation Date} & \colhead{Configuration} & \colhead{Baselines} \\
\colhead{} & \colhead{(UT)} & \colhead{} & \colhead{(m)}}
\startdata
IRAS 04016+2610 & Sep. 3 2012, Jan. 22 2013, Mar. 17, 19, 19, Oct. 3, Dec. 29 2014 & E, B, C, C, C, E, D & 4$\,$-$\,$982 \\
IRAS 04108+2803B & Mar. 20, Jun. 19, Oct. 3 2014, Jan. 15 2015 & C, E, E, D & 6$\,$-$\,$386 \\
IRAS 04158+2805 & Sep. 3 2012, Jan. 21, Feb. 1 2013, Mar. 20, Oct. 3, Dec. 30 2015 & E, B, B, C, E, D & 5$\,$-$\,$982 \\
IRAS 04166+2706 & Oct. 3 2012, Jan. 2, 21, Feb 1 2013 & E, C, B, B & 5$\,$-$\,$982 \\
IRAS 04169+2702 & Oct. 3 2012, Jan. 2, 22 2013 & E, C, B & 5$\,$-$\,$982 \\
IRAS 04181+2654A & Oct. 5 2014, Jan. 2 2015 & E, D & 7$\,$-$\,$152 \\
IRAS 04181+2654B & Mar. 17, 19, 19, Oct. 3 2014, Jan. 3 2015 & C, C, C, E, D & 7$\,$-$\,$386 \\
IRAS 04263+2426 & Mar. 17, 19, 19, Jun. 19, Oct. 3, Dec. 29 2014 & C, C, C, E, E, D & 5$\,$-$\,$386 \\
IRAS 04295+2251 & Oct. 4 2014, Jan. 2 2015 & E, D & 7$\,$-$\,$153 \\
IRAS 04302+2247 & Sep. 3 2012, Jan. 2 2013, Jan. 3 2015  & E, C, D & 1$\,$-$\,$386 \\
IRAS 04365+2535 & Sep. 3 2012, Jan. 15 2015 & E, D & 5$\,$-$\,$152 \\
\enddata
\end{deluxetable}

The masses of protoplanetary disks are an important driver for the processes of star and planet formation. Early in the lifetime of protostars disks are thought to be massive and turbulent, and accretion from the envelope onto these massive disks could cause gravitational instabilities that drive high accretion rates in young sources \citep[e.g.][]{Kenyon1987}. The disk mass also sets a limit on the amount of material available for forming planets and the ultimate outcomes of the planet formation process \citep[e.g.][]{Alibert2005}.

Disk masses are typically measured from their sub-millimeter flux, which if tracing optically thin matter, is directly proportional to the amount of material present in the disk \citep[e.g.][]{Beckwith1990}. Class II disks are the easiest to study because, without a protostellar envelope, the entirety of the sub-millimeter flux can be attributed to disk emission. In the past decade there has been a large effort, particularly with interferometers like CARMA, the SMA, and now ALMA, towards measuring Class II disk masses \citep{Andrews2005,Andrews2007,Eisner2008,Mann2010,Mann2014,Ansdell2016,Barenfeld2016,Pascucci2016,Ansdell2017}. These studies typically find that the majority of these disks fall well below the $0.01-0.1$ M$_{\odot}$ needed to form planetary systems like our own \citep[e.g.][]{Weidenschilling1977,Desch2007}.

It may be that by the typical age of Class II disks \citep[$1-5$ Myr;][]{Andre1994,Barsony1994}, dust grain growth has locked up large amounts of mass in large bodies to which sub-millimeter observations are not sensitive. If this is the case, then studying the disks around the younger \citep[$\sim0.5$ Myr;][]{Evans2009} Class I disks, which have had less significant dust processing, may give a better picture of the initial mass budget for forming planets. The masses of these disks are more difficult to determine because they are still embedded in their natal envelope, and any millimeter flux measurement will include a contribution from both the disk and envelope. Masses for Class I disks have been measured from high resolution millimeter visibilities by using radiative transfer modeling to separate disk and envelope contributions \citep{Eisner2005,Eisner2012,Sheehan2014}, but sample sizes for these surveys are small.

In this paper we present a study of a sample of 10 Class I protostars in the Taurus Molecular Cloud, expanding on our previous work by including new, high resolution CARMA 1.3 mm maps for an expanded sample of objects. We use radiative transfer modeling and employ a fitting method that uses Markov Chain Monte Carlo simulations to fit models simultaneously to a 1.3 mm visibilities + broadband SED dataset and measure physical properties of the systems such as disk masses and radii. We discuss how these measurements of Class I disk masses compare to measurements of Class II disk masses, and what this means for the formation of planets.

\section{Observations \& Data Reduction}
\label{section:observations}

\subsection{Sample Selection}

\begin{figure*}[t]
\centering
\figurenum{1a}
\includegraphics[width=7in]{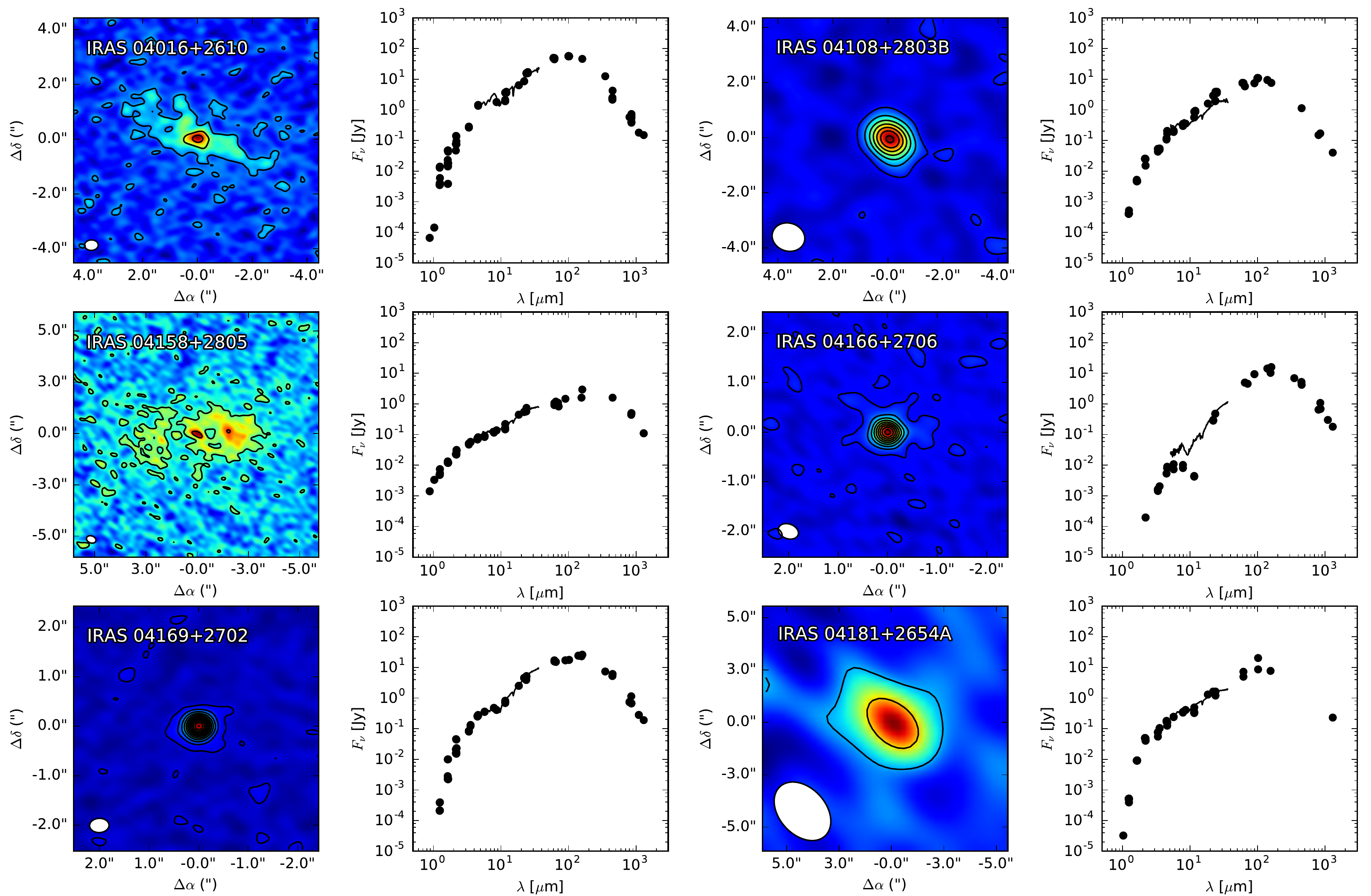}
\caption{We show the 1.3 mm CARMA maps (first and third columns) and broadband SEDs (second and fourth columns) for each of the sources in our sample. Contours start at $2\sigma$ and subsequent contours are at intervals of 5$\sigma$. For these maps, $\sigma$ ranges from 0.5 -- 1.5 mJy. Many of our sources were observed with high enough spatial resolution to resolve structure in their disks and envelopes. Only one source, I04181B is undetected in our maps. For all sources the SED is sampled across the electromagnetic spectrum and includes a high resolution Spitzer IRS spectrum.}
\label{fig:carma_sources1}
\end{figure*}

Our sample includes 10 protostars in Taurus that are consistently identified as Class I across multiple independent studies \citep[e.g.][]{Myers1987,Kenyon1993a,Motte2001,Andrews2005,Furlan2008,Eisner2012}. All of our targets fit standard criteria for selecting Class I protostars: all have an infrared spectral index of $\alpha > 0.15$ and a bolometric temperature of $70 < T_{bol} < 650$ \citep[e.g.][]{Myers1987,Chen1995,Motte2001,Andrews2005}. Furthermore, all of our targets have been observed with the Spitzer IRS spectrograph and most have silicate and/or CO$_2$ ice absorption in their spectra, commonly associated with embedded sources \citep[e.g.][]{Alexander2003,Watson2004,Boogert2004,Pontoppidan2008}. 

In this sample we have excluded Class I objects that have been identified as compact binaries because the modeling of close separation binaries can be more challenging \citep[e.g.][]{Sheehan2014}. In all, our sample contains 10 of 12 companionless bona fide Class I protostars in Taurus. We were unable to observe the remaining 2, which were left for last because they were expected to be faint, before CARMA was decommissioned. Our targets do, however, span a wide range of millimeter fluxes \citep[][]{Motte2001,Jorgensen2009,Eisner2012} so they should span a range of masses of circumstellar material. They also span a range of spectral types \citep[M6-K4;][]{White2004,Doppmann2005,Connelley2010} and scattered light morphologies \citep[e.g.][]{Padgett1999,Stark2006,Gramajo2010}.

\subsection{CARMA 1.3 mm Observations}

We obtained 230 GHz Combined Array for Research in Millimeter-wave Astronomy (CARMA) dust continuum observations of our sample from September 3, 2012 until January 15, 2015. The observations were taken with CARMA's B, C, D, and E configurations (baselines ranging from $\sim$5 m to $\sim$1 km) so that our data would be sensitive to both large and small scale structures from the protostellar disks and envelopes. The observations were set up with 14 of CARMA's 16 spectral windows in wideband continuum mode from 216.798 GHz to 233.296 GHz with 500 MHz of bandwidth per-spectral window. The continuum observations had a mean frequency of 222.242 GHz and a total of 7 GHz of continuum bandwidth. The remaining two spectral windows were configured for spectral line observations, which we will discuss in a separate paper. We show a log of our observations in Table \ref{table:obs}.

\begin{figure*}[t]
\centering
\figurenum{1b}
\includegraphics[width=7in]{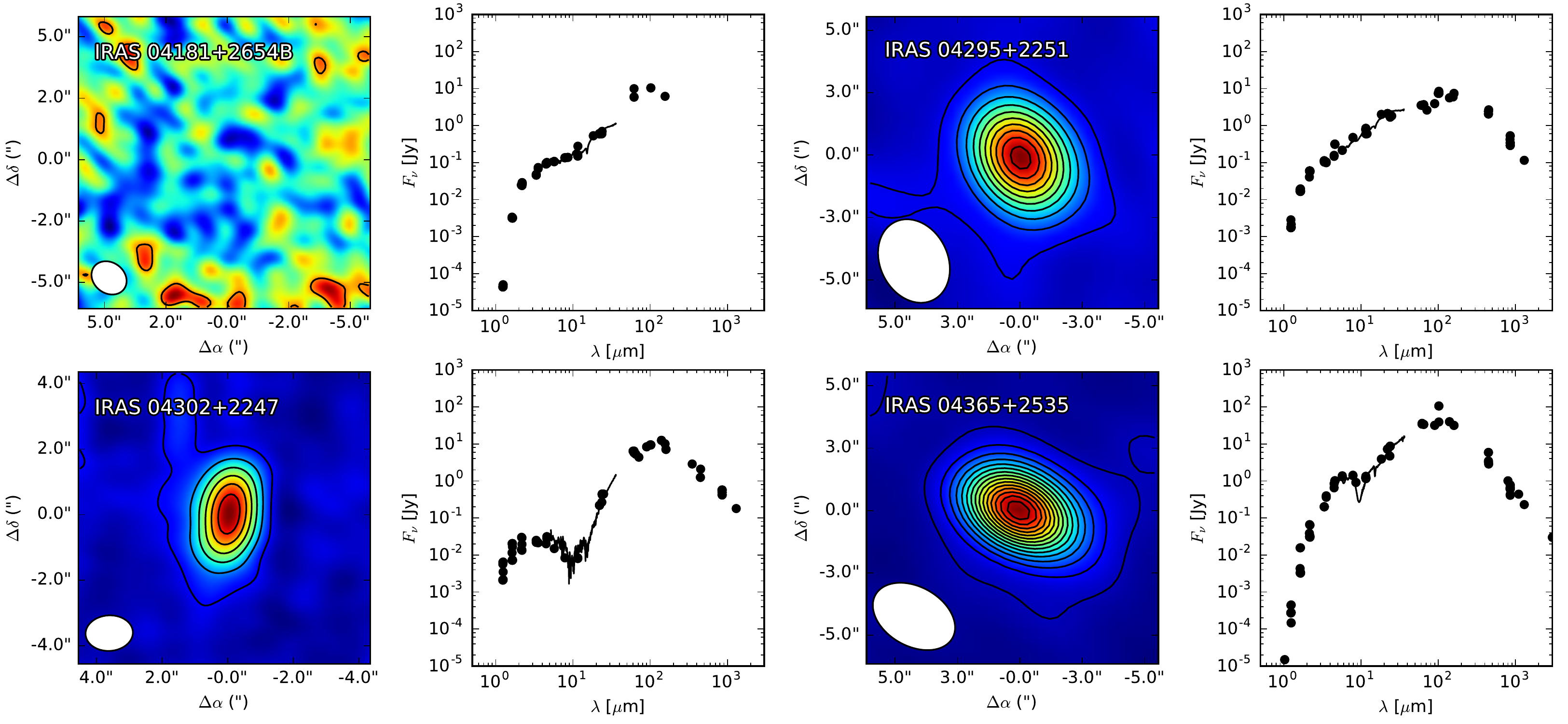}
\caption{Continued.}
\label{fig:carma_sources2}
\end{figure*}

The CARMA data were reduced using the \texttt{CASA} software package in the standard way. For the majority of the tracks Uranus was used as the flux calibrator, the quasar 3C84 as the bandpass calibrator, and 3C111 and QSO 0510+180 as the gain calibrators. For a few tracks, QSO 0530+135 was also used as the gain calibrator when 3C111 or QSO 0510+180 were unavailable. For tracks where Uranus was unavailable to use as the flux calibrator we used 3C84 instead with measured fluxes from the SMA calibrator catalog.

Following calibration, the data were imaged by Fourier transforming the visibilities with \texttt{CASA}'s {\it clean} routine to produce images of our targets. For each source we combine all of the available tracks and configurations to produce a single image. We use the multi-frequency synthesis mode and the Briggs weighting scheme with a robust parameter of 0.5. Because CARMA is a heterogeneous array, mosaicking mode is needed to correctly image the data. We show images of our targets in Figures \ref{fig:carma_sources1} \& \ref{fig:carma_sources2}. Although we show images of the data, we do all of our analysis and modeling directly with the visibilities.

\subsection{SEDs from the Literature}

For each of our sources we compiled a broadband SED using data from the literature. This data includes photometry from Spitzer IRAC and MIPS, WISE, 2MASS, and IRAS as well as other infrared and millimeter surveys \citep{Ladd1991,Barsony1992,MoriartySchieven1994,Ohashi1996,Chandler2000,Motte2001,Young2003,Andrews2005,Andrews2007,Eisner2012}. In addition to this photometry, we downloaded a calibrated Spitzer IRS spectrum with wavelength coverage from $5-30$ $\mu$m from the CASSIS database to include in our SED \citep{Lebouteiller2011,Lebouteiller2015}.

In order to assess the quality of our model fits through metrics such as $\chi^2$, which we describe in Section \ref{section:fitting}, we assume a uniform 10\% flux uncertainty on all photometry from the literature. We also sample the IRS spectrum at 25 points spaced uniformly over the spectral range to include in our SED. We do this because calculating fluxes at the several hundred IRS spectrum channels with our radiative transfer modeling routines is computationally expensive, and the goal of this paper is not to model in extreme detail the IRS spectrum.

\subsection{HST Scattered Light Images}

Five of our sources (IRAS 04016+2610, IRAS 04108+2803B, IRAS 04158+2805, IRAS 04295+2251 and IRAS 0302+2247) have near-infrared Hubble Space Telescope scattered light images with the Wide-field Planetary Camera available, although IRAS 04108+2803 is a non-detection. We downloaded calibrated versions of these images from the Hubble Legacy Archive for comparison with our models.

\section{Modeling}
\label{section:modeling}

We use detailed radiative transfer modeling to produce synthetic observations of a protostar model that can be matched to our millimeter visibilities + broadband SED dataset. The model includes a central star, protoplanetary disk, and a rotating collapsing envelope, following the modeling scheme of \citet{Eisner2005}, \citet{Eisner2012}, and \citet{Sheehan2014}. These previous studies ran large grids of radiative transfer models and fit those grids to multi-wavelength datasets to determine system parameters. The availability of computational resources, however, limited those previous studies to a small set of discrete values for each parameter. Here we have developed a Markov Chain Monte Carlo procedure to more completely explore parameter space, particularly in the vicinity of the best fit model. We describe the components and free parameters of the model as well as our modeling technique below.

\subsection{Pre-Main-Sequence Star}

Our Class I model includes a central protostar with a temperature of 4000 K and a luminosity, $L_{*}$, that left as a free parameter. The majority of the sources in our system are K- or M-type stars \citep{White2004,Doppmann2005,Connelley2010}, so a temperature of 4000 K is a reasonable assumption. We may, however, explore varying the protostellar temperature in future works.

\subsection{Disk}

Our model also includes a protoplanetary disk that uses the standard density profile of a flared power-law disk,
\begin{equation}
\rho = \rho_0 \left(\frac{R}{R_0}\right)^{-\alpha} \, \exp\left(-\frac{1}{2}\left[\frac{z}{h(R)}\right]^2\right),
\end{equation}
where $R$ and $z$ are in cylindrical coordinates. $h(R)$ is the disk scale height at a given radius,
\begin{equation}
h(R) = h_0 \left(\frac{R}{1 \text{ AU}}\right)^{\beta}.
\end{equation}
The surface density profile is
\begin{equation}
\Sigma = \Sigma_0 \left(\frac{R}{R_0}\right)^{-\gamma}, \, \gamma = \alpha - \beta.
\end{equation}

We truncate the disk at a specified inner and outer disk radius, $R_{in}$ and $R_{disk}$, that are allowed to vary in our fit. The surface density power law exponent, $\gamma$, and the scale height power law exponent, $\beta$, are also left as free parameters in our model. We leave the disk mass, $M_{disk}$, and scale height at 1 AU, $h_0$, as free parameters. The density at the inner radius, $\rho_0$ can be calculated from the disk mass by integrating equation 1 over all space.

\subsection{Envelope}

Our sources are young and likely embedded in an envelope of material remaining from the initial cloud from which they formed, so we also include an envelope component in our protostar model. We use the density profile for a rotating collapsing envelope from \citet{Ulrich1976},
\begin{equation}
\footnotesize
\rho = \frac{\dot{M}}{4\pi}\left(G M_* r^3\right)^{-\frac{1}{2}} \left(1+\frac{\mu}{\mu_0} \right)^{-\frac{1}{2}} \left(\frac{\mu}{\mu_0}+2\mu_0^2\frac{R_c}{r}\right)^{-1}
\end{equation}
where $\mu = \cos \theta$, and $r$ and $\theta$ are defined in the typical sense for spherical coordinates. We truncate the envelope at the same inner radius, $R_{in}$, as the disk and at an outer radius, $R_{env}$, that is left as a free parameter. We require that the envelope radius be larger than the disk radius. $R_c$ is the critical radius, inside of which the envelope begins to flatten due to rotation, and is the location where the majority of material is accreting onto the disk \citep{Ulrich1976,Terebey1984}. This makes the most sense physically if the critical radius is equal to the disk radius, so in our model we specify that $R_c = R_{disk}$. The envelope mass, $M_{env}$, is also a free parameter, and the density normalization can again be calculated by integrating equation 4 over all space. 

We give the envelope an outflow cavity. In regions where
\begin{equation}
z > 1 \text{ AU} + r^{\zeta}
\end{equation}
we reduce the envelope density by the factor $f_{cav}$. We leave both $\zeta$ and $f_{cav}$ as free parameters to be varied in our modeling routines.

\subsection{Dust}

We provide our disk model with dust opacities that are the same as those used by \citet{Sheehan2014}, that for small maximum dust grain sizes, are similar to the icy dust grains from \citet{Ossenkopf1994}. The opacities have a composition that is 40\% astronomical silicate, 30\% organics, and 30\% water ice, roughly following the recipe from \citet{Pollack1994} but adjusted to match the dense protostellar core opacities from \citet{Ossenkopf1994} (see \citealt{Sheehan2014} for a more thorough discussion). We use a gain size distribution with $n \propto a^{-p}$ with $p = 3.5$ \citep{Mathis1977}, and dust grains ranging from 0.005 $\mu$m to $a_{max}$. In the envelope, where dust grain growth is likely to be less advanced, we fix $a_{max} = 1$ $\mu$m. In the disk, however, we leave $a_{max}$ as a free parameter.

\subsection{Radiative Transfer Modeling + Synthetic Images}

\floattable
\begin{deluxetable}{lccccccccccccccc}
\rotate
\tablecaption{Best-fit Model Parameters}
\tablenum{3}
\tabletypesize{\tiny}
\label{table:best_fits}
\tablehead{\colhead{Source} & \colhead{$L_{*}$} & \colhead{$M_{disk}$} & \colhead{$R_{in}$} & \colhead{$R_{disk}$} & \colhead{$h_0$} & \colhead{$\gamma$} & \colhead{$\beta$} & \colhead{$M_{env}$} & \colhead{$R_{env}$} & \colhead{$f_{cav}$} & \colhead{$\zeta$} & \colhead{$a_{max}$} & \colhead{$i$} & \colhead{$P.A.$} & \colhead{$A_K$}\\ \colhead{ } & \colhead{$[L_{\odot}]$} & \colhead{$[M_{\odot}]$} & \colhead{[AU]} & \colhead{[AU]} & \colhead{[AU]} & \colhead{} & \colhead{} & \colhead{$[M_{\odot}]$} & \colhead{[AU]} & \colhead{} & \colhead{} & \colhead{$[\mu$m]} & \colhead{$[\,^{\circ}\,]$} & \colhead{$[\,^{\circ}\,]$} & \colhead{}}
\startdata
IRAS 04016+2610 & $ 7.0^{+ 0.9}_{- 0.7}$ & $ 0.009^{+ 0.001}_{- 0.001}$ & $  1.2^{+  0.3}_{-  0.3}$ & $  497^{+   19}_{-   22}$ & $0.12^{+0.03}_{-0.03}$ & $0.66^{+0.03}_{-0.03}$ & $1.03^{+0.06}_{-0.07}$ & $ 0.023^{+ 0.010}_{- 0.004}$ & $ 1446^{+  156}_{-  109}$ & $0.59^{+0.26}_{-0.28}$ & $1.00^{+0.30}_{-0.33}$ & $  1095^{+   721}_{-   395}$ & $  67^{+   1}_{-   1}$ & $   64^{+    1}_{-    1}$ & $1.1^{+0.2}_{-0.3}$ \\
IRAS 04108+2803B & $ 0.4^{+ 0.1}_{- 0.1}$ & $ 0.015^{+ 0.005}_{- 0.005}$ & $  0.4^{+  0.1}_{-  0.1}$ & $   46^{+   14}_{-   13}$ & $0.10^{+0.01}_{-0.01}$ & $1.27^{+0.33}_{-0.41}$ & $0.89^{+0.10}_{-0.10}$ & $ 0.004^{+ 0.003}_{- 0.002}$ & $  369^{+  281}_{-  142}$ & $0.49^{+0.08}_{-0.11}$ & $0.76^{+0.19}_{-0.12}$ & $    70^{+   115}_{-    42}$ & $  37^{+   6}_{-   7}$ & $  102^{+   43}_{-   36}$ & \nodata \\
IRAS 04158+2805 & $ 0.5^{+ 0.0}_{- 0.0}$ & $ 0.148^{+ 0.028}_{- 0.027}$ & $  0.1^{+  0.0}_{-  0.0}$ & $  571^{+   12}_{-   12}$ & $0.21^{+0.01}_{-0.01}$ & $-0.36^{+0.04}_{-0.03}$ & $0.88^{+0.02}_{-0.02}$ & $ 0.078^{+ 0.107}_{- 0.046}$ & $ 3823^{+ 3338}_{- 1916}$ & $0.00^{+0.00}_{-0.00}$ & $0.91^{+0.01}_{-0.01}$ & $   542^{+   114}_{-    73}$ & $  65^{+   1}_{-   1}$ & $   93^{+    1}_{-    1}$ & \nodata \\
IRAS 04166+2706 & $ 0.2^{+ 0.0}_{- 0.0}$ & $ 0.027^{+ 0.003}_{- 0.003}$ & $  0.6^{+  0.3}_{-  0.1}$ & $  180^{+    7}_{-    6}$ & $0.04^{+0.01}_{-0.01}$ & $1.96^{+0.02}_{-0.02}$ & $0.78^{+0.09}_{-0.08}$ & $ 0.100^{+ 0.009}_{- 0.009}$ & $ 1209^{+   64}_{-   48}$ & $0.64^{+0.08}_{-0.06}$ & $1.01^{+0.03}_{-0.03}$ & $ 11151^{+  2487}_{-  2449}$ & $  35^{+   2}_{-   1}$ & $  151^{+    2}_{-    2}$ & \nodata \\
IRAS 04169+2702 & $ 0.8^{+ 0.1}_{- 0.1}$ & $ 0.012^{+ 0.001}_{- 0.001}$ & $  0.1^{+  0.0}_{-  0.0}$ & $   39^{+    1}_{-    1}$ & $0.03^{+0.01}_{-0.01}$ & $0.71^{+0.04}_{-0.05}$ & $1.03^{+0.06}_{-0.05}$ & $ 0.055^{+ 0.004}_{- 0.005}$ & $  672^{+   39}_{-   41}$ & $0.05^{+0.01}_{-0.01}$ & $1.02^{+0.01}_{-0.01}$ & $ 11904^{+  2571}_{-  1853}$ & $  41^{+   1}_{-   2}$ & $    0^{+    2}_{-    2}$ & \nodata \\
IRAS 04181+2654A & $ 0.3^{+ 0.0}_{- 0.0}$ & $ 0.006^{+ 0.001}_{- 0.001}$ & $  0.1^{+  0.1}_{-  0.0}$ & $   47^{+   22}_{-   14}$ & $0.17^{+0.04}_{-0.05}$ & $0.12^{+0.40}_{-0.37}$ & $0.85^{+0.11}_{-0.16}$ & $ 1.234^{+ 0.688}_{- 0.391}$ & $22853^{+ 9014}_{- 5557}$ & $0.36^{+0.17}_{-0.13}$ & $1.21^{+0.17}_{-0.14}$ & $     3^{+     7}_{-     2}$ & $   7^{+  10}_{-   5}$ & $   92^{+   59}_{-   53}$ & \nodata \\
IRAS 04181+2654B & $ 0.1^{+ 0.0}_{- 0.0}$ & $ 0.000^{+ 0.000}_{- 0.000}$ & $  0.1^{+  0.0}_{-  0.0}$ & $   24^{+    4}_{-    4}$ & $0.10^{+0.01}_{-0.01}$ & $0.73^{+0.12}_{-0.13}$ & $0.60^{+0.04}_{-0.04}$ & $ 0.005^{+ 0.008}_{- 0.002}$ & $  570^{+  567}_{-  206}$ & $0.47^{+0.05}_{-0.04}$ & $1.45^{+0.03}_{-0.05}$ & $    93^{+    43}_{-    34}$ & $   3^{+   2}_{-   1}$ & $   91^{+   23}_{-   26}$ & \nodata \\
IRAS 04295+2251 & $ 0.3^{+ 0.0}_{- 0.0}$ & $ 0.018^{+ 0.001}_{- 0.001}$ & $  0.2^{+  0.0}_{-  0.0}$ & $  127^{+   11}_{-    9}$ & $0.13^{+0.01}_{-0.01}$ & $0.53^{+0.12}_{-0.14}$ & $1.11^{+0.04}_{-0.03}$ & $ 0.037^{+ 0.008}_{- 0.006}$ & $ 1081^{+  132}_{-  114}$ & $0.22^{+0.04}_{-0.04}$ & $0.89^{+0.04}_{-0.06}$ & $  1621^{+   342}_{-   294}$ & $  29^{+   5}_{-   5}$ & $   67^{+    5}_{-    5}$ & \nodata \\
IRAS 04302+2247 & $ 0.4^{+ 0.0}_{- 0.0}$ & $ 0.114^{+ 0.026}_{- 0.019}$ & $  3.1^{+  0.5}_{-  0.4}$ & $  244^{+    6}_{-    6}$ & $0.05^{+0.04}_{-0.02}$ & $-0.30^{+0.10}_{-0.12}$ & $0.77^{+0.13}_{-0.15}$ & $ 0.017^{+ 0.006}_{- 0.004}$ & $ 1086^{+  260}_{-  196}$ & $0.69^{+0.24}_{-0.36}$ & $1.26^{+0.17}_{-0.28}$ & $ 31174^{+ 27954}_{- 12633}$ & $  76^{+   1}_{-   1}$ & $  172^{+    1}_{-    1}$ & \nodata \\
IRAS 04365+2535 & $ 2.1^{+ 0.2}_{- 0.2}$ & $ 0.030^{+ 0.002}_{- 0.003}$ & $  0.3^{+  0.1}_{-  0.1}$ & $  143^{+   19}_{-   18}$ & $0.07^{+0.02}_{-0.01}$ & $0.98^{+0.17}_{-0.24}$ & $0.90^{+0.09}_{-0.10}$ & $ 0.071^{+ 0.035}_{- 0.019}$ & $ 1829^{+  192}_{-  175}$ & $0.49^{+0.14}_{-0.17}$ & $0.86^{+0.05}_{-0.04}$ & $  1949^{+   886}_{-   374}$ & $  54^{+   2}_{-   5}$ & $   76^{+    4}_{-    4}$ & $1.3^{+0.1}_{-0.2}$ \\
\enddata
\end{deluxetable}

We use the 3D Monte Carlo radiative transfer modeling codes \texttt{RADMC-3D} \citep{Dullemond2012} and \texttt{Hyperion} \citep{Robitaille2011} to produce synthetic observations of our protostar model that can be subsequently be compared with our combined millimeter visibilities + broadband SED dataset. We use the radiative transfer codes to run a simulation to calculate the temperature everywhere throughout the disk and envelope by propagating photon packets through the model and updating the temperature in each model cell every time a photon is absorbed and then reemitted. 

In most cases we use \texttt{RADMC-3D} to do the temperature calculation, however for protostars with a particularly high density, i.e. small disk or envelope radii or large disk or envelope masses, we use \texttt{Hyperion} because it can be run in parallel to speed up the computation. We have compared the results from \texttt{RADMC-3D} and \texttt{Hyperion} when running the same input model and find that the codes are consistent. Following the radiative transfer simulation we use raytracing in \texttt{RADMC-3D} to produce synthetic SEDs and millimeter images, and we Fourier transform the millimeter images to produce synthetic visibilities. 

The viewing angle parameters, inclination and position angle ($i$ and $p.a.$), are free parameters in our fitting procedure. Rather than include positional offsets as free parameters, we use simple Gaussian fits to the data to determine offsets for each source and to center the data.

\subsection{Fitting Procedure}
\label{section:fitting}

We fit our model to the data by comparing synthetic visibilities and SEDs to our millimeter visibilities + broadband SED dataset with the Markov Chain Monte Carlo (MCMC) code \texttt{emcee} \citep{ForemanMackey2013}. \texttt{emcee} uses an implementation of the Goodman \& Weare affine invariant ensemble sampler to explore parameter space. We assume uniform priors on each of the parameters, with the following limits: $0.1\text{\,L}_{\odot} \leq L_* \leq 10 \text{\,L}_{\odot}$, $M_{disk} < 0.3$ M$_{\odot}$, $0.1\text{\,AU} < R_{in} < R_{disk} < R_{env}$, $0.01\text{\,AU} < h_0$, $-0.5 < \gamma < 2$, $0 < \beta < 2$, $M_{env} < 1$ M$_{\odot}$, $0 < f_{cav} < 1$, $0.5 < \zeta < 1.5$, $1\,\mu\text{m} < a_{max} < 1 \text{\,cm}$, $0^{\circ} < i < 90^{\circ}$, $0^{\circ} < p.a. < 180^{\circ}$, and $0 < A_K < 2$. For each source in our sample we run a MCMC fit in which we spread out 200 walkers randomly with a uniform distribution over a large volume of parameter space and the walkers collectively move towards regions of parameters space that represent better fits to the data. 


In these simulations the walkers are seeking to maximize the log-likihood of the model, which is directly proportional to $\chi^2$. Here we are simultaneously fitting to the millimeter visibilities and the broadband SED, which are separate datasets with heteroscedastic error bars, so specifying a goodness-of-fit metric is challenging. For simplicity we use the weighted sum of the $\chi^2$ values for our individual datasets,
\begin{equation}
X^2 = w_{vis} \, \chi_{vis}^2 + w_{SED} \, \chi_{SED}^2,
\end{equation}
to provide a log-likihood to our fits, and we seek to maximize $-X^2/2$. $\chi_{vis}^2$ is calculated by directly comparing the real and imaginary components of the data and model visibilities, although in subsequent figures we show the one-dimensional azimuthally averaged visibility amplitudes and millimeter images as they are easier to interpret. We can vary the weights of each dataset ($w_*$) to increase the contribution of that dataset to the fit. As resolved images provide more direct information about source geometry than unresolved SEDs, we typically weight up the visibilities to ensure that they are fit well. We have found empirically \citep[e.g. see][]{Sheehan2014} that using $w_{vis} = 10$ and $w_{SED} = 1$ provide a good balance in fitting both datasets simultaneously. We use those values for fitting all of our sources.

\begin{figure*}
\centering
\figurenum{2a}
\includegraphics[width=7in]{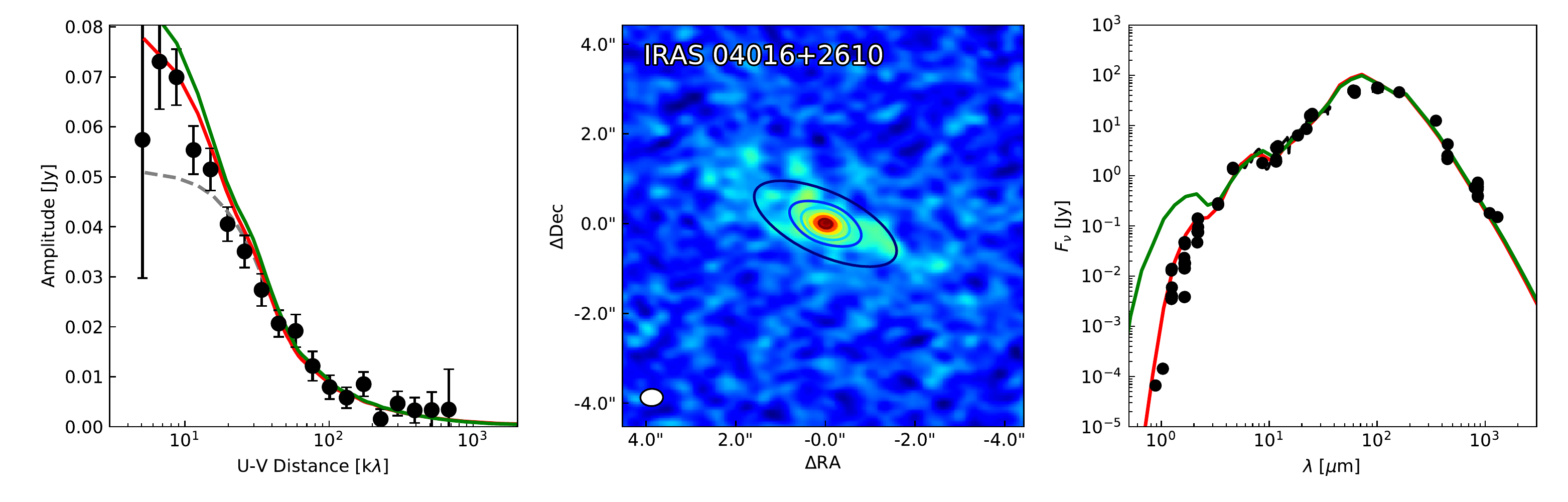}
\caption{We show the 1.3 mm visibility profile ({\it left}), 1.3 mm image ({\it center}), and broadband SED ({\it right}) for IRAS 04016+2610 with the best-fit disk+envelope model curves over-plotted in green on the left and right and as contours in the center panel. We also show the disk contribution to the visibilities as a gray dashed line. The green curve shows our base model, which matches the visibilities but does not extinct the spectrum sufficiently at short wavelengths. If we include some foreground extinction in the fit (the red line), however, the models fit the data. Parameters for these models can be found in Table \ref{table:best_fits}.}
\label{fig:fit1}
\end{figure*}

We allow the walkers to explore parameter space for an extended burn-in period, during which the walkers converge towards regions of parameter space that fit the data well. We consider the walkers to have converged once there is a negligible change in the parameter values compared with the spread of the walkers over a large number of steps. This criterion is satisfied for most targets, but may not be formally satisfied for some parameters that are not well constrained by our datasets for a few objects. Waiting for those parameters to formally converge would take too long due to the high computational demand of this procedure. The generation of a radiative transfer model for a single set of parameter values (i.e. a single step of a single walker) is computationally intensive to run and can take anywhere from a few minutes to a few hours. \texttt{emcee} uses MPI to spread the calculations out over a large number of cores, with each core computing the models for a subset of walkers, to significantly speed up the computation. In principle the calculations can be spread over any number of nodes, but we find that fits typically converge over reasonable timescales of a few weeks when run using a single node with 28 cpus. We can run fits to many sources simultaneously, each on a single node of a supercomputer.

\section{Results}
\label{section:results}

\begin{figure*}[t]
\centering
\figurenum{2b}
\includegraphics[width=7in]{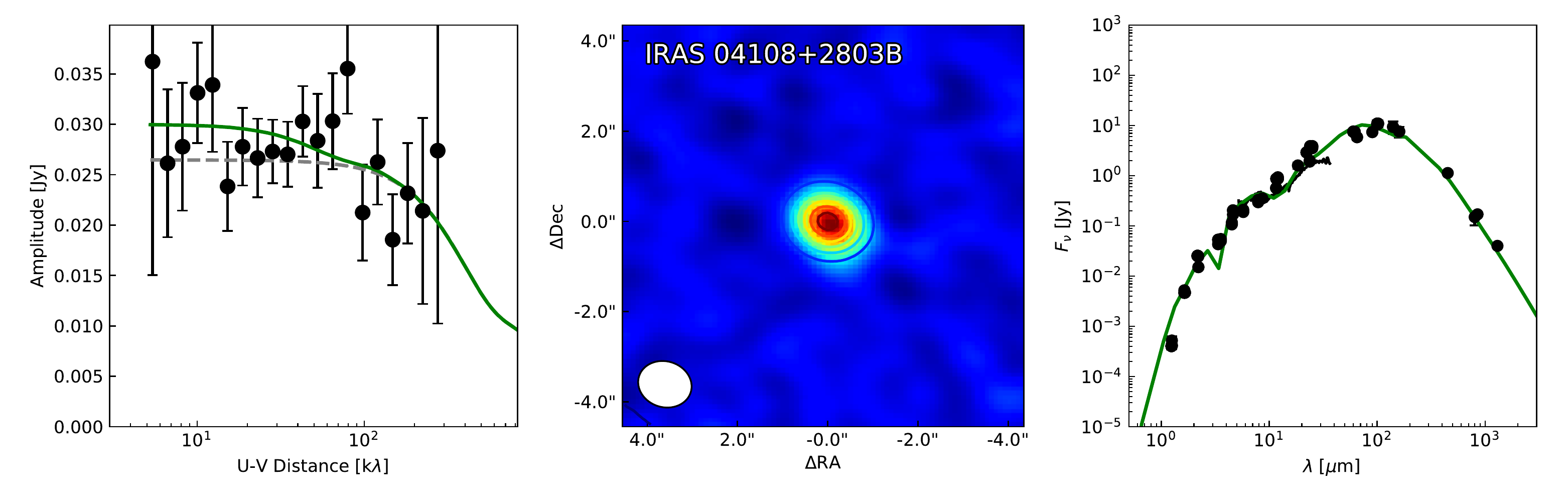}
\caption{We show the 1.3 mm visibility profile ({\it left}), 1.3 mm image ({\it center}), and broadband SED ({\it right}) for IRAS 04108+2803B with the best-fit disk+envelope model curves over-plotted in green on the left and right and as contours in the center panel. We also show the disk contribution to the visibilities as a gray dashed line. Parameters for these models can be found in Table \ref{table:best_fits}.}
\label{fig:fit2}
\end{figure*}

\begin{figure*}[t]
\centering
\figurenum{2c}
\includegraphics[width=7in]{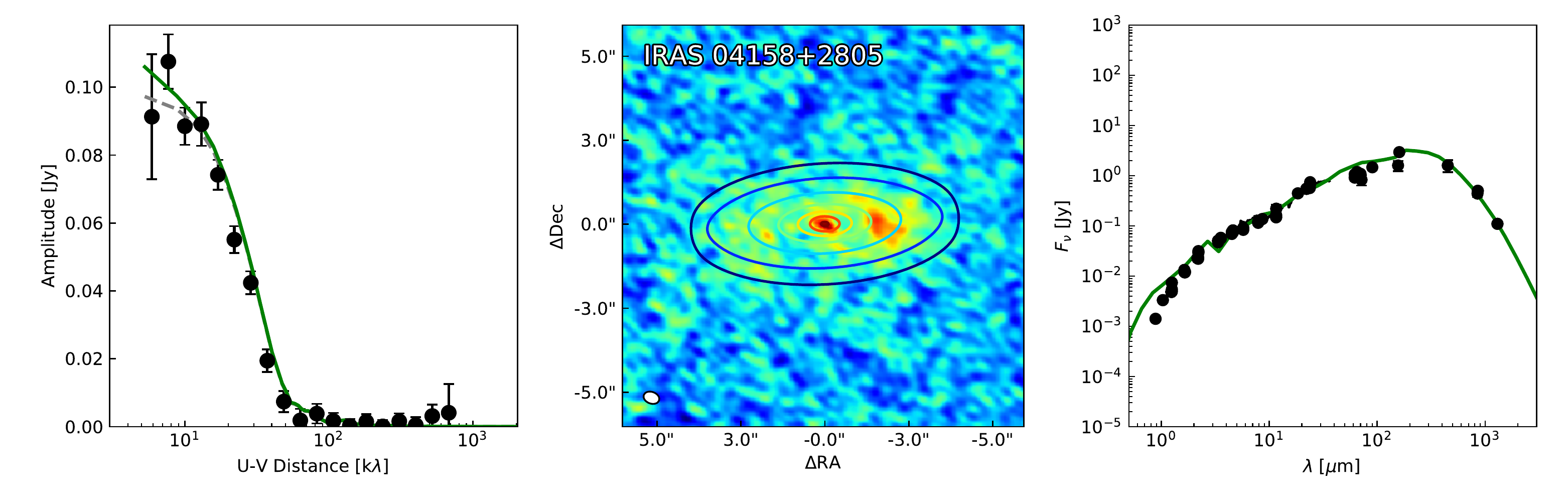}
\caption{We show the 1.3 mm visibility profile ({\it left}), 1.3 mm image ({\it center}), and broadband SED ({\it right}) for IRAS 04158+2805 with the best-fit disk+envelope model curves over-plotted in green on the left and right and as contours in the center panel. We also show the disk contribution to the visibilities as a gray dashed line. Parameters for these models can be found in Table \ref{table:best_fits}.}
\label{fig:fit3}
\end{figure*}

\begin{figure*}[t]
\centering
\figurenum{2d}
\includegraphics[width=7in]{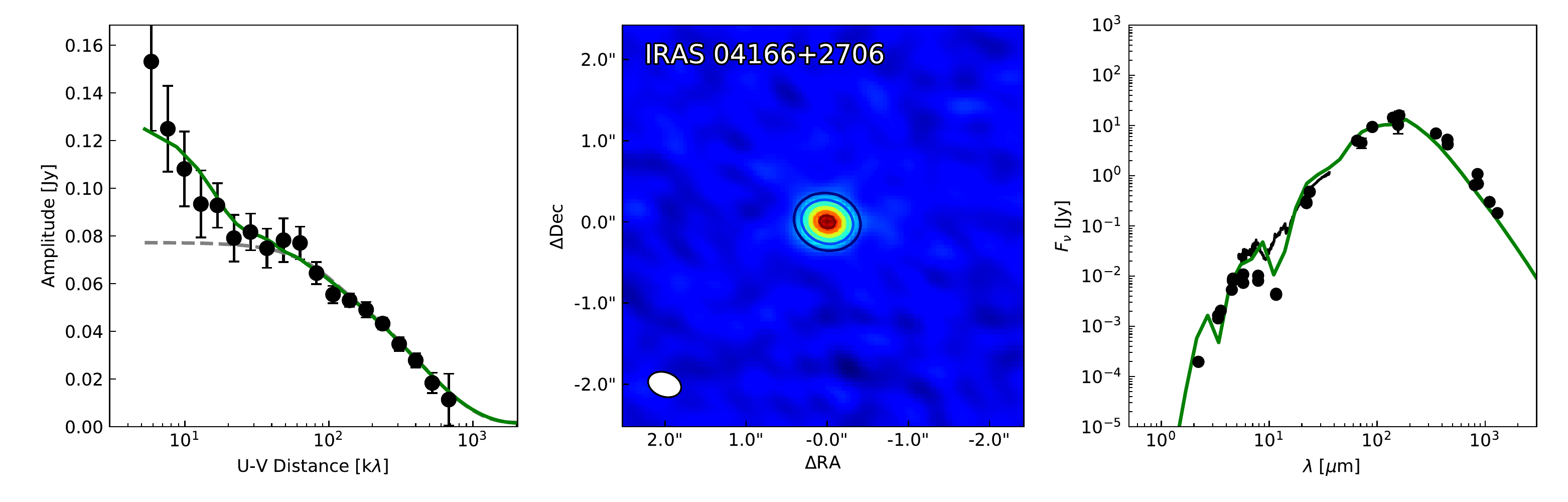}
\caption{We show the 1.3 mm visibility profile ({\it left}), 1.3 mm image ({\it center}), and broadband SED ({\it right}) for IRAS 04166+2706 with the best-fit disk+envelope model curves over-plotted in green on the left and right and as contours in the center panel. We also show the disk contribution to the visibilities as a gray dashed line. Parameters for these models can be found in Table \ref{table:best_fits}.}
\label{fig:fit4}
\end{figure*}

\begin{figure*}[t]
\centering
\figurenum{2e}
\includegraphics[width=7in]{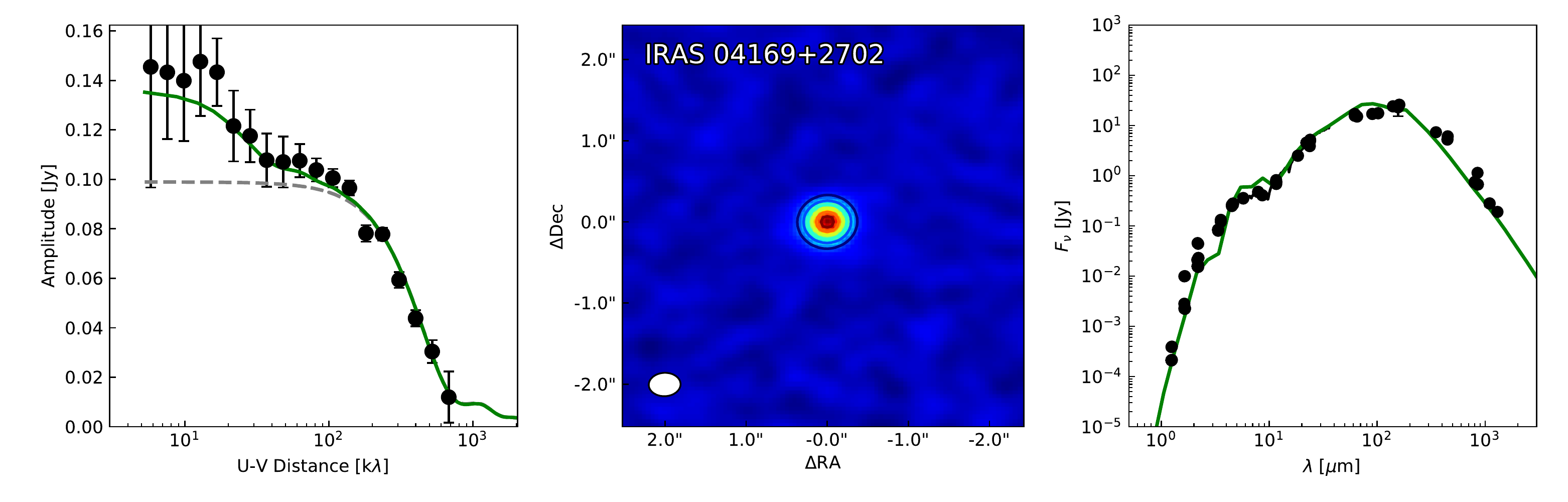}
\caption{We show the 1.3 mm visibility profile ({\it left}), 1.3 mm image ({\it center}), and broadband SED ({\it right}) for IRAS 04169+2702 with the best-fit disk+envelope model curves over-plotted in green on the left and right and as contours in the center panel. We also show the disk contribution to the visibilities as a gray dashed line. Parameters for these models can be found in Table \ref{table:best_fits}.}
\label{fig:fit5}
\end{figure*}

\begin{figure*}[t]
\centering
\figurenum{2f}
\includegraphics[width=7in]{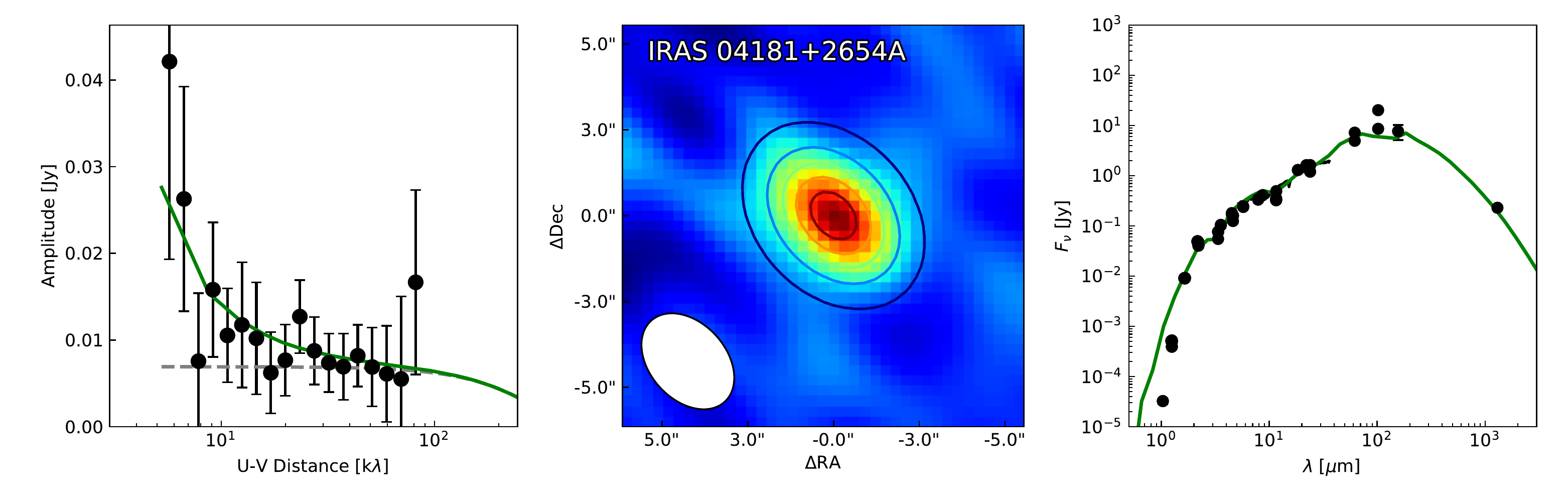}
\caption{We show the 1.3 mm visibility profile ({\it left}), 1.3 mm image ({\it center}), and broadband SED ({\it right}) for IRAS 04181+2654A with the best-fit disk+envelope model curves over-plotted in green on the left and right and as contours in the center panel. We also show the disk contribution to the visibilities as a gray dashed line. Parameters for these models can be found in Table \ref{table:best_fits}.}
\label{fig:fit6}
\end{figure*}

\begin{figure*}[t]
\centering
\figurenum{2g}
\includegraphics[width=7in]{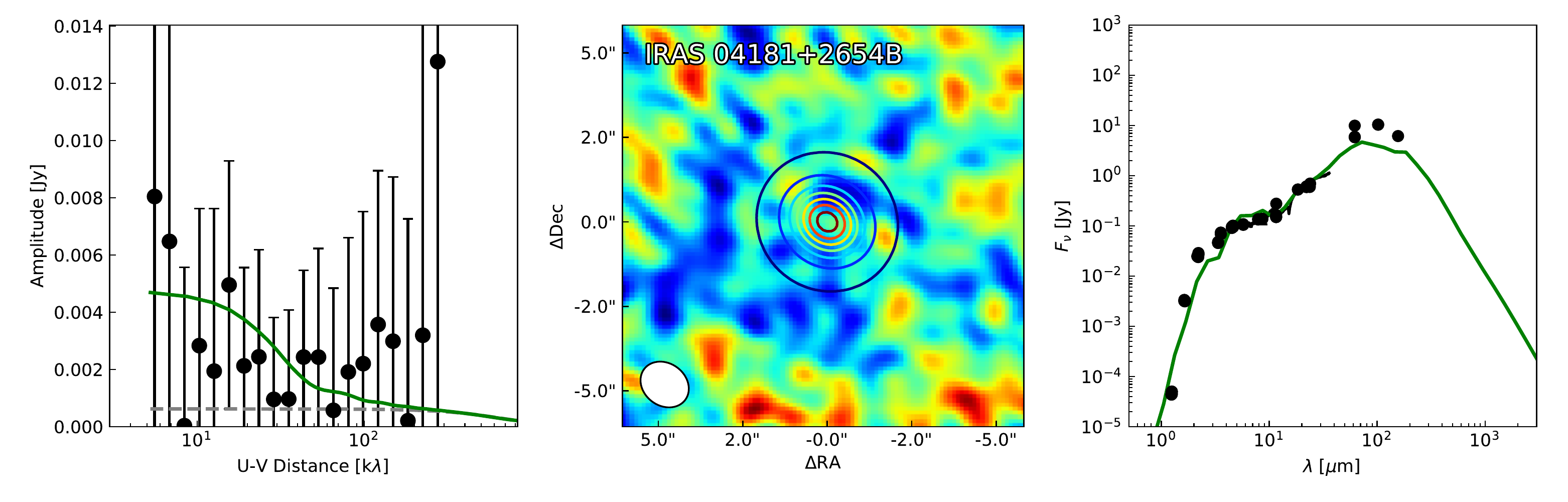}
\caption{We show the 1.3 mm visibility profile ({\it left}), 1.3 mm image ({\it center}), and broadband SED ({\it right}) for IRAS 04181+2654B with the best-fit disk+envelope model curves over-plotted in green on the left and right and as contours in the center panel. We also show the disk contribution to the visibilities as a gray dashed line. Parameters for these models can be found in Table \ref{table:best_fits}.}
\label{fig:fit7}
\end{figure*}

\begin{figure*}[t]
\centering
\figurenum{2h}
\includegraphics[width=7in]{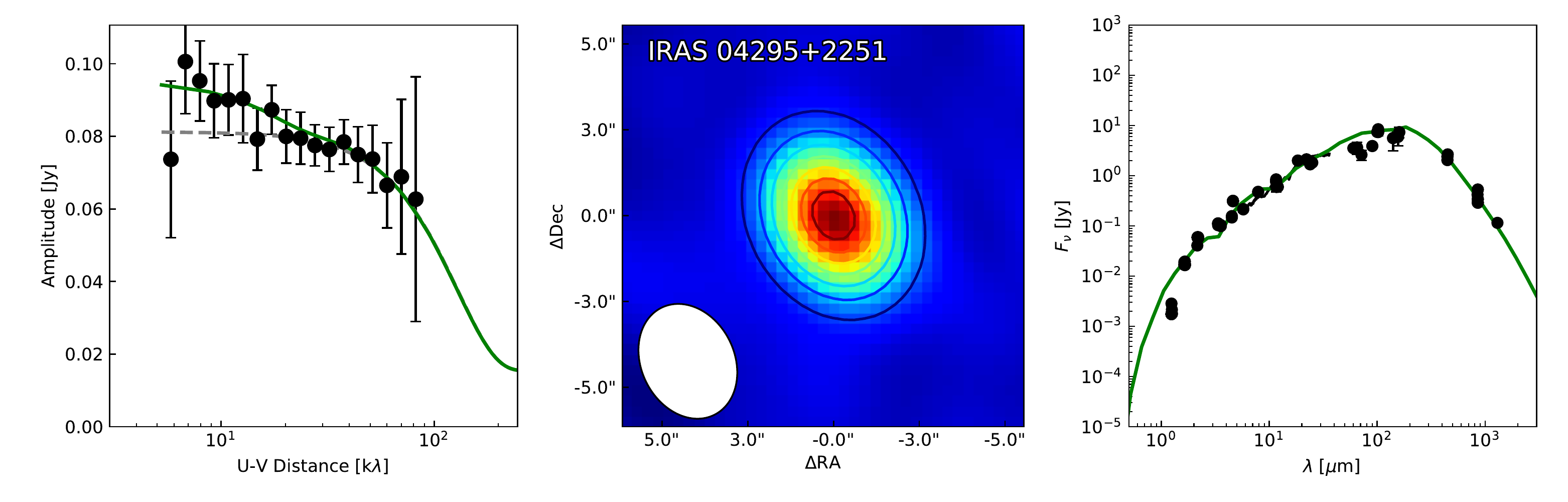}
\caption{We show the 1.3 mm visibility profile ({\it left}), 1.3 mm image ({\it center}), and broadband SED ({\it right}) for IRAS 04295+2251 with the best-fit disk+envelope model curves over-plotted in green on the left and right and as contours in the center panel. We also show the disk contribution to the visibilities as a gray dashed line. Parameters for these models can be found in Table \ref{table:best_fits}.}
\label{fig:fit8}
\end{figure*}

\begin{figure*}[t]
\centering
\figurenum{2i}
\includegraphics[width=7in]{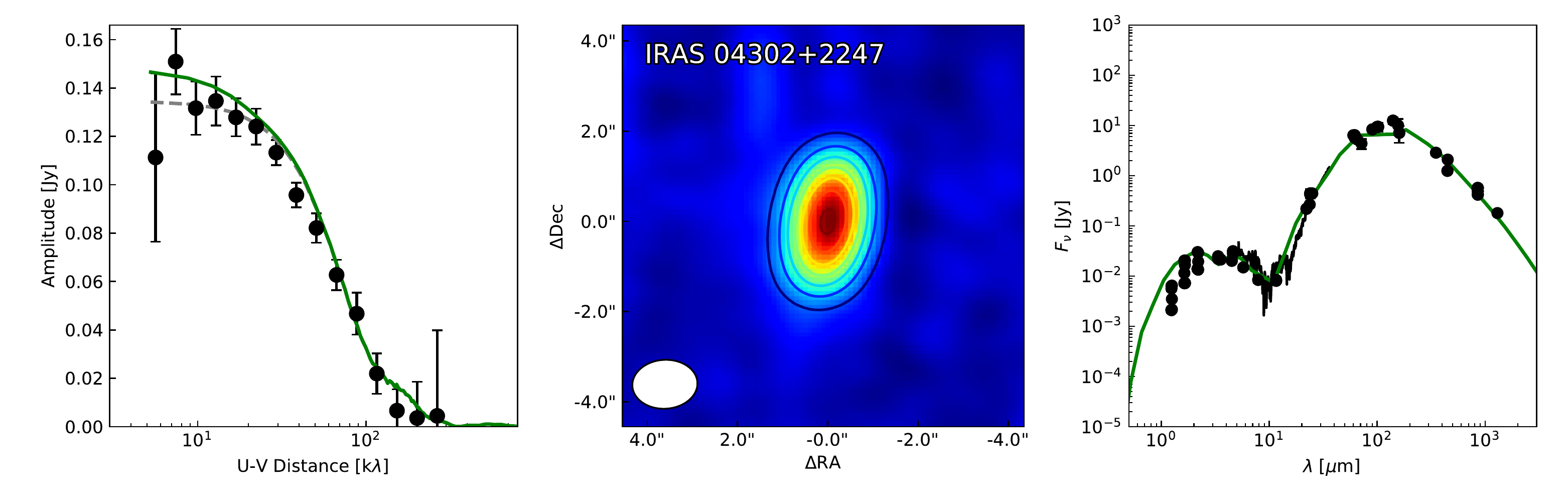}
\caption{We show the 1.3 mm visibility profile ({\it left}), 1.3 mm image ({\it center}), and broadband SED ({\it right}) for IRAS 04302+2247 with the best-fit disk+envelope model curves over-plotted in green on the left and right and as contours in the center panel. We also show the disk contribution to the visibilities as a gray dashed line. Parameters for these models can be found in Table \ref{table:best_fits}.}
\label{fig:fit9}
\end{figure*}

\begin{figure*}[t]
\centering
\figurenum{2j}
\includegraphics[width=7in]{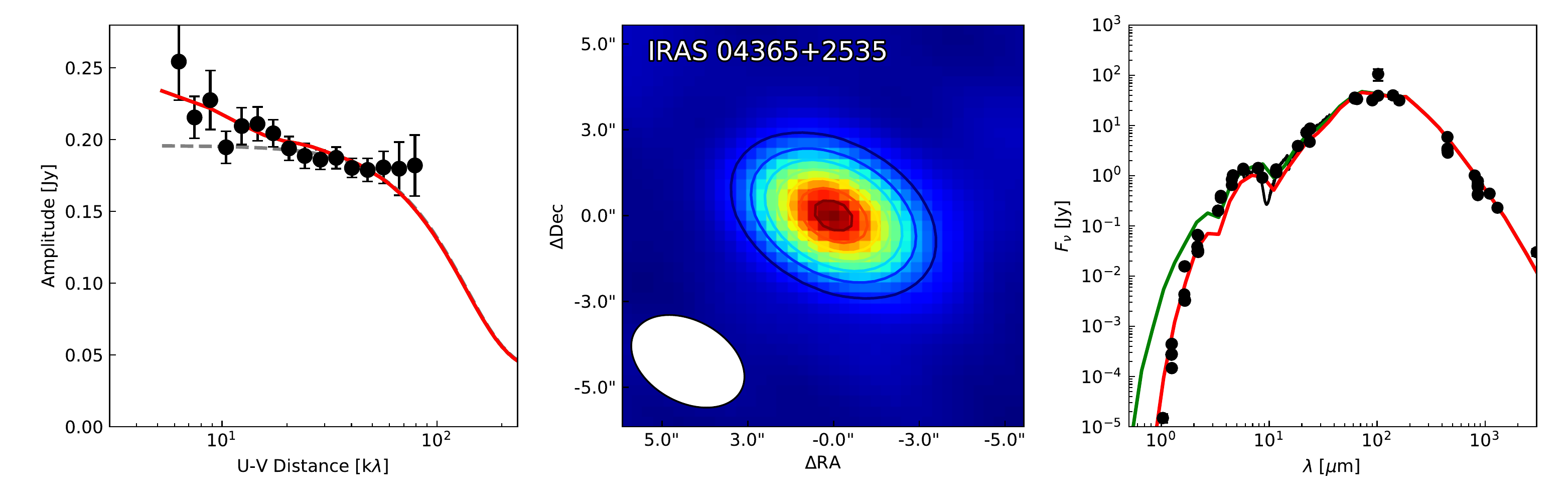}
\caption{We show the 1.3 mm visibility profile ({\it left}), 1.3 mm image ({\it center}), and broadband SED ({\it right}) for IRAS 04365+2535 with the best-fit disk+envelope model curves over-plotted in green on the left and right and as contours in the center panel. We also show the disk contribution to the visibilities as a gray dashed line. The green curve shows our base model, which matches the visibilities but does not extinct the spectrum sufficiently at short wavelengths. If we include some foreground extinction in the fit (the red line), however, the models fit the data. Parameters for these models can be found in Table \ref{table:best_fits}.}
\label{fig:fit10}
\end{figure*}

Our model fitting procedure is able to find models that reproduce the combined 1.3 mm visibilities + broadband SED dataset for each of our sources. We list the best fit parameters for these models in Table \ref{table:best_fits} and show the best fit models compared with the data for each source in Figures \ref{fig:fit1}-\ref{fig:fit10}. We note that the masses (both disk and envelope) listed here assume a standard gas-to-dust ratio of 100. Dust masses, which are the values that are directly constrained by our modeling, are a factor of 100 lower. We list total mass for ease of comparing with the Minimum Mass Solar Nebula, which is typically quoted in terms of total mass.

The best fit parameters in Table \ref{table:best_fits} are the median values of the posterior distribution for each parameter, after burn-in is discarded. We also derive uncertainties on the listed values as the range around the median that contains 68\% of the posterior distribution. While these uncertainties are a reasonable representation of the range of allowed values for each parameter, our weighted sum of $\chi^2$ likely makes it such that these are not rigorous uncertainties. We have, however, compared the uncertainties we measure on inclination with the results of a simple uniform disk geometrical fit and find that the magnitudes of the errors are generally in agreement. As such, the errors we list are likely reasonable estimates of how well constrained our models are.

\begin{figure*}[b!]
\centering
\figurenum{5a}
\includegraphics[width=7in]{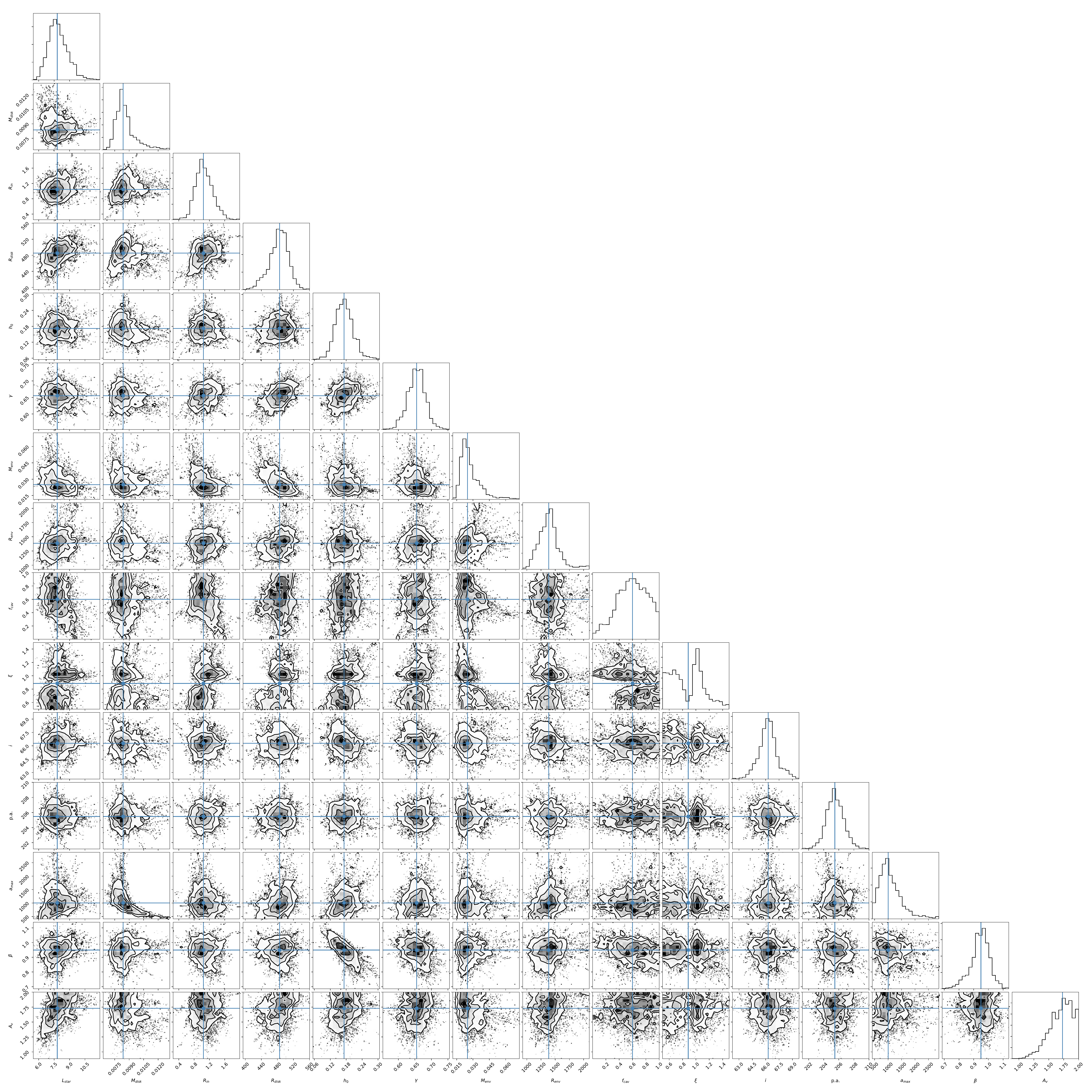}
\caption{Projected posterior pdf for every combination of parameter pairs for IRAS 04016+2610.}
\label{fig:triangle_I04016}
\end{figure*}

\begin{figure*}
\centering
\figurenum{5j}
\includegraphics[width=7in]{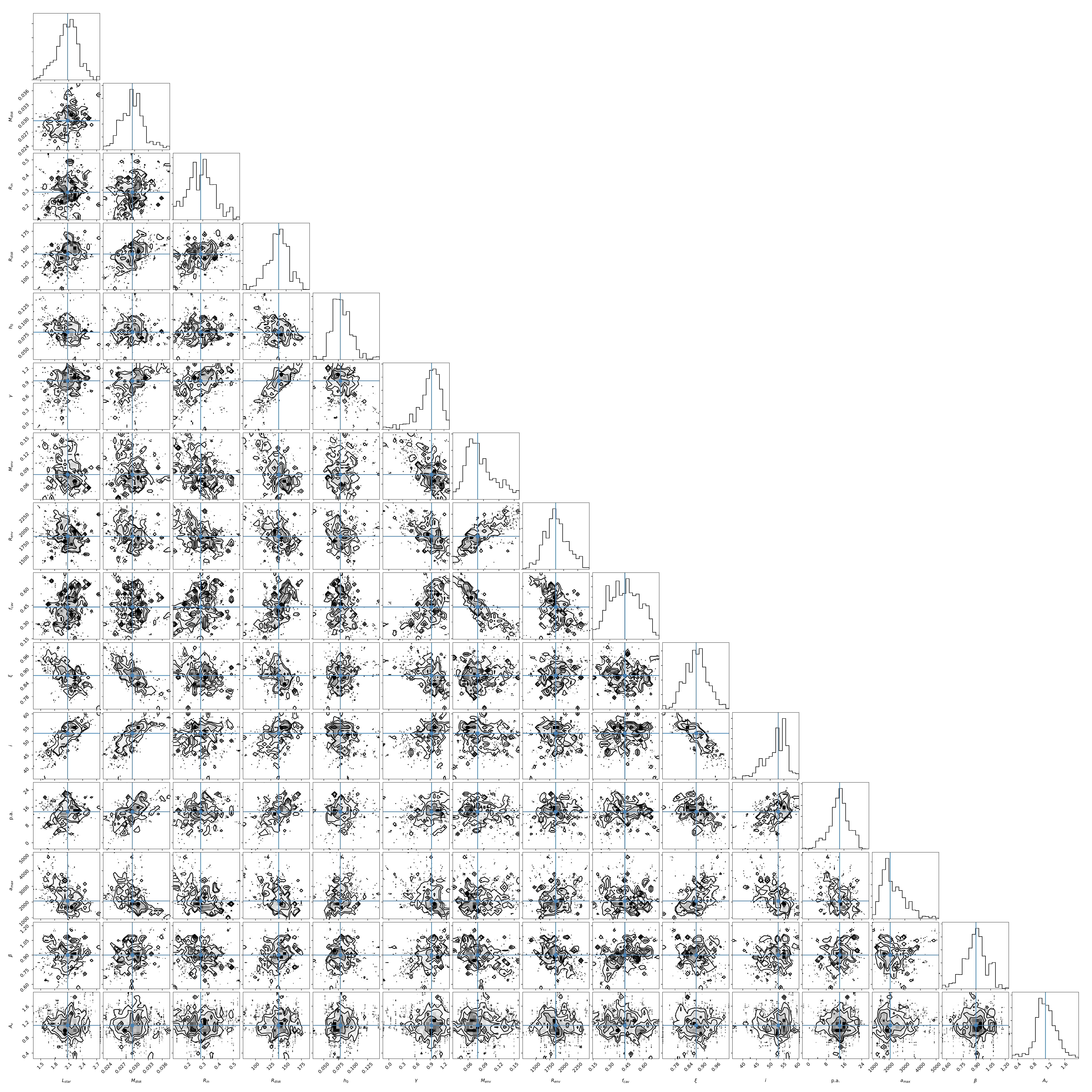}
\caption{Projected posterior pdf for every combination of parameter pairs for IRAS 04365+2535.}
\label{fig:triangle_I04365}
\end{figure*}

We show triangle plots of the projected posterior distributions for all parameter pairs for each source in Figures \ref{fig:triangle_I04016} -- \ref{fig:triangle_I04365}. These plots are useful for assessing the range of parameters allowed by our fits, and also for determining what degeneracies might be present. For example, in our fits we find that envelope mass and radius ($M_{env}$ and $R_{env}$) are often degenerate, likely because for many of our sources we lack the short baselines in our millimeter data needed to fully probe envelope structure.

Our sample has a range of inferred properties, including disk radii ranging from $25 - 570$ AU and disk masses ranging from $0.0002 - 0.15$ M$_{\odot}$. Our sample also has a diversity of envelope properties, with masses ranging from $0.003 - 1$ M$_{\odot}$ and radii from $400 - 10000$ AU. The ratio of disk-to-envelope masses ranges from $0.2 - 5$. We discuss each of the sources below.

\subsection{IRAS 04016+2610}

IRAS 04016+2610 has one of the largest disks in our sample, with a radius of about 500 AU. For such a large disk, though, it is relatively low mass, at 0.009 M$_{\odot}$. The disk is highly inclined, with an inclination of 67$^{\circ}$. Although we did not include scattered light imaging in our fit, our best fit model nicely reproduces the observed HST scattered light image of the system (see Figure \ref{fig:scattered_light}). The envelope is about twice as massive as the disk, indicating that IRAS 04016+2610 is a well-embedded source.

Our base model is not able to fully reproduce both the millimeter visibilities and the SED for IRAS 04016+2610 simultaneously. Any fit that reproduces the millimeter visibilities does not provide enough extinction to match the SED at near-infrared wavelengths (see Figure \ref{fig:fit1}), so some additional source of extinction is needed. To remedy this, we have run a fit that includes an additional parameter, the K-band extinction ($A_K$) that we use to redden the SED using the \citet{McClure2010} extinction law, and find that both datasets can be reproduced with $A_K \sim 1.5$. 

Although this extinction could simply be from the large scale cloud in the foreground of IRAS 04016+2610, previous studies of the system have suggested other possibilities. \citet{Hogerheijde2000} found that IRAS 04016+2610 is in close proximity to a neighboring starless dark cloud, and \citet{Brinch2007a} found that they could only fit their models if IRAS 04016+2610 was located behind the edge of that dark cloud. If this dark cloud is indeed in the foreground, as \citet{Brinch2007a} suggests, it could be the source of the extinction. Alternatively, it may be that this large amount of extinction could come from large scale, constant density material from the cloud that has not yet begun to collapse, but could collapse sometime in the future \citep[e.g.][]{Jayawardhana2001}.

IRAS 04016+2610 was previously studied using a similar procedure to our own modeling by \citet{Eisner2012}, but using a grid rather than an MCMC fit. The parameters for the best fits IRAS 04016+2610 are similar to what we find here, with a typical disk mass of 0.005 M$_{\odot}$ and a disk radius of 250 - 450 AU. Most of the best fit models from \citet{Eisner2012} for IRAS 04016+2610, however, are found to have $i\sim35-40^{\circ}$, much smaller than what we find here. The exception to this a model in which the scattered light image is given more weight, and as a result the best fit inclination is 65$^{\circ}$. This is also consistent with the inclination \citet{Stark2006} found, of $i\sim65^{\circ}$ by modeling only the near-infrared scattered light image. Measurements of the inclination from the bipolar outflow found to be associated with IRAS 04016+2610 \citep{Gomez1997,Hogerheijde1998} find that the disk must have an inclination of 60$^{\circ}$, in very good agreement with what we find here.

Other studies have previously modeled this source and found a range of results. \citet{Furlan2008} found a much lower inclination ($i\sim40^{\circ}$) and disk radius ($R_c\sim100$ AU), but only considered the SED and had no imaging constraints on the system geometry from imaging. Similarly, \citet{Robitaille2007} found low inclinations from a SED-only fit. \citet{Gramajo2010} find a higher inclination, of $50-63^{\circ}$ by considering the Spitzer IRS spectrum and scattered light images, along with the broadband SED. \citet{Brinch2007a} found that the IRAS 04016+2610 has a slightly flattened envelope with an inclination of $74^{\circ}$, while \citet{Brinch2007b} suggested that the disk may be misaligned with the envelope and has an inclination of $40^{\circ}$, but these models were based on lower resolution observations than we present here. Inferred disk masses for this source range from $\sim0.004-0.02$ M$_{\odot}$, and our measurement falls nicely in the middle of that range.

\subsection{IRAS 04108+2803B}

Our best fit model for IRAS 04108+2803B has both a compact disk ($R_{disk} \approx 50$ AU) and envelope ($R_{env} \approx 370$ AU), and the disk is about three times as massive as the envelope. The disk is not resolved well in our millimeter maps, nor is the system detected in scattered light, so the constraints on geometrical properties of the system are somewhat weak.

This system has been modeled previously and found to have a compact disk, with a disk radius of 30-100 AU and moderate (20-60$^{\circ}$) inclinations \citep{Kenyon1993a,Whitney1997,Eisner2005,Furlan2008}. Our best fit model is in good agreement with \citet{Eisner2005}, who find a disk radius of 30 AU, an envelope radius of 500 AU, and an inclination of 24$^{\circ}$. They find that the disk is significantly more massive than our results ($M_{disk}\sim0.5$ M$_{\odot}$), but they also suggest that this is likely an overestimate.

\citet{Chiang1999} suggested that the SED of this source could be fit by an inclined flared accretion disk, suggesting that the disk may be an edge-on Class II disk rather than a Class I source. The large disk radius needed ($\sim250$ AU), though, would have been resolved in our millimeter observations, and indeed \citet{Eisner2005} find that an envelope component is needed to fit the SED. \citet{Watson2004} also suggest that the 15.2 $\mu$m ice absorption feature found in the Spitzer IRS spectrum is most likely to arise in an envelope. This is consistent with our own results that find that an envelope is needed to match the data. 

Our results do, however, show that the envelope is quite low-mass compared to other Class I sources, which seems to suggest that IRAS 04108+2803B is close to dispelling its envelope and emerging as a Class II system. This is consistent with the presence of a wide-separation companion, IRAS 04108+2803A, that appears to be a more evolved, Class II system. If the binary system is approximately coeval, as might be expected, then these sources may both be young and on the boundary between Class I and II.

\subsection{IRAS 04158+2805}

IRAS 04158+2805 has the largest disk of the sample, at $R_{disk} = 570$ AU, and is the most massive disk (M$_{disk} = 0.15$ M$_{\odot}$). The disk is somewhat inclined, at about 65$^{\circ}$. The envelope has a mass of $M_{env} = 0.08$ M$_{\odot}$ and a radius of $R_{env} = 3800$ AU. Like IRAS 04016+2610, we did not include the HST scattered light image in the fit, but our best fit model naturally reproduces the scattered light image without any fitting needed (see Figure \ref{fig:scattered_light}).

There has been some disagreement about the nature of this object in previous studies. Most signs point to this source being a very low mass protostar, with a spectral type of M5-6 ($M_* \sim 0.1-0.2$) \citep{White2004,Luhman2006,Connelley2010}, although mass estimates from gas kinematics \citep{Andrews2008} and other spectral typing surveys \citep{Doppmann2005} have suggested it might be more massive. Some studies have classified IRAS 04158+2805 as a Class II disk, and indeed \citet{Glauser2008} suggested that the near-infrared scattered light image and SED for the system could be fit without an envelope component. However, their model needs a much larger disk radius ($R_{disk} \sim 1150$ AU) than what we find here. Our observations suggest that the disk is much smaller than that, although still quite large compared to typical protoplanetary disks. Moreover, the infrared spectrum exhibits absorption features of H$_2$O and CO$_2$ ices and a silicate absorption feature, all of which are more commonly associated with Class I sources embedded in envelopes \citep[e.g.][]{Watson2004,Pontoppidan2008}. The presence of these features along with the good fit of our disk+envelope model to the combined SED and millimeter visibilities suggest that this is an embedded source.

\subsection{IRAS 04166+2706}

Our best fit model for IRAS 04166+2706 indicates a 180 AU radius disk and an envelope that is about four times more massive than its disk. The disk and the envelope are clearly detected in our millimeter visibilities, with an apparent break at 30-80 k$\lambda$ where the disk begins to dominate over the envelope. There is no apparent flattening at the shortest baselines, likely indicating that we have resolved out some of the envelope, and may be underestimating its mass. No HST scattered light image was available for the source, and \citet{Eisner2005} were unable to detect it in scattered light with Keck LRIS imaging. This is perhaps unsurprising, given how embedded the source appears to be from the SED.

IRAS 04166+2706's defining characteristic is it's bipolar outflow \citep{Bontemps1996} that has an extremely high velocity component that is highly collimated \citep{Tafalla2004,SantiagoGarcia2009,Wang2014}. That, coupled with its highly embedded disk, have led some to suggest that it is a Class 0 protostar. \citet{Tafalla2004} suggested based on the outflow that the disk must be highly inclined, although our high resolution millimeter observations contradict that.

The disk mass of our best fit model for IRAS 04166+2706 is in good agreement with the results from \citet{Eisner2012}, but the disk radius we measure is much smaller (180 AU compared with 450 AU). \citet{Furlan2008} find a disk radius of 300 AU, although note that a disk of 200 AU can also provide a good fit. \citet{Kenyon1993a} find a smaller disk (70 AU), but a similar inclination (30$^{\circ}$). Our millimeter dataset, however is much higher resolution than what was available for \citet{Eisner2012}, and \citet{Kenyon1993a} and \citet{Furlan2008} model only the SED, so we are able to constrain the structure of the disk.

\subsection{IRAS 04169+2702}

IRAS 04169+2702 has a compact ($R_{disk} \sim 40$ AU) disk hidden in a larger envelope that is about four times more massive than the disk. The disk is has a mass of $M_{disk} \sim 0.012$, and it is being viewed at a moderate inclination of $\sim41^{\circ}$. The millimeter visibility amplitudes flatten out at around 50 k$\lambda$, likely where the disk begins to dominate over the envelope.

This source was modeled previously by \citet{Eisner2012}, who found a much larger disk, typically 250 -- 450 AU although weighting up the SED produces a fit with a 100 AU disk, but comparable disk masses and inclinations. \citet{Furlan2008} fit the SED with a disk about twice the size we find here, but with a high inclination, while \citet{Robitaille2007} found from SED fitting that $R_{disk}<150$ AU and $i>30^{\circ}$, both in agreement with our results. IRAS 04169+2702 is associated with a bipolar outflow \citep{Bontemps1996}, and \citet{Ohashi1997} find that the outflow is associated with an elongated envelope structure inclined $60^{\circ}$ with respect to our line of sight \citep{Ohashi1997}. However, compared with both of these studies we have much better resolution to study disk structure, so our measurement is likely more accurate.

\subsection{IRAS 04181+2654A}

IRAS 04181+2654A appears to be a low mass disk ($M_{disk}\sim0.006$ M$_{\odot}$) embedded in a very massive envelope ($M_{env}\sim1$ M$_{\odot}$). Although the visibilities are noisy, a clear break in the visibility profile at around 10 k$\lambda$ is readily identifiable, indicating the presence of significant amounts of emission on large spatial scales. Our observations are not sensitive to large enough scales to fully determine the structure of the envelope, but it appears to be quite large ($R_{env}\sim20000$ AU) and massive. The disk, by comparison, is quite compact, with a radius of about 50 AU and a mass of only 0.006 $M_{\odot}$.

Because this object has few flux measurements at millimeter wavelengths, there have been a lack of studies to determine parameters for the system, and what has been done only considered the SED. Our results are in good agreement with what was found by \citet{Furlan2008}, who find a low inclination disk with a radius of 50 AU and an envelope with a radius of 10,000 AU. \citet{Kenyon1993a} also find that the disk is compact ($R_{disk} = 70$ AU) and low inclination ($i = 30^{\circ}$).

\subsection{IRAS 04181+2654B}

IRAS 04181+2654B is detected in the near- to far-infrared, but has not been detected at millimeter wavelengths. This remains true of our own observations, which detect no 1.3 mm emission. It seems to be embedded based on CO$_2$ ice absorption in its Spitzer IRS SED and its association with the embedded source IRAS 04181+2654A. We have included the source in our modeling, but the models are not constrained well. We show that the disk is likely small and low mass, but can say little else definitively. At 31" from IRAS 04181+2654A, or 4300 AU projected separation, it falls well within the envelope we measure for IRAS 04181+2654A. As that envelope is quite large and massive, it could be that this source is simply a low mass disk hidden behind the IRAS 04181+2654A envelope.

\subsection{IRAS 04295+2251}

\begin{figure*}
\centering
\figurenum{3}
\includegraphics[width=7in]{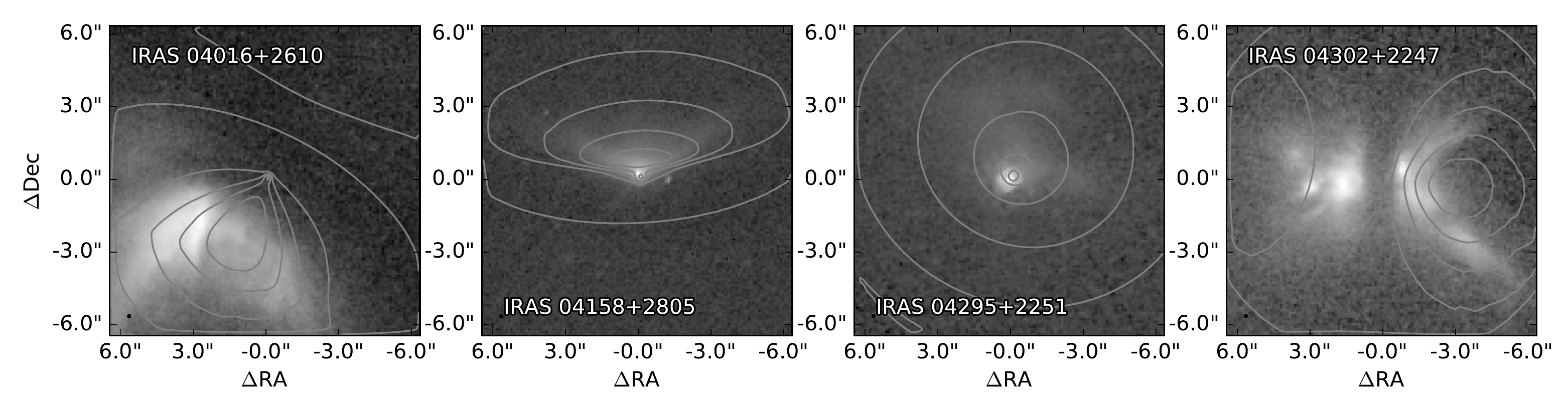}
\caption{0.8 $\mu$m scattered light images from HST for the four sources where such images were available. We show an 0.8 $\mu$m scattered light image produced from the best fit model as grey contours. In all four cases, although we did not fit our model to the scattered light data, the best fit model does a reasonable job of reproducing the the scattered light distribution. IRAS 04032+2247 shows a more edge-on morphology than we find when fitting the combined millimeter visibilities and broadband SED dataset, possibly indicating a disk warp or disk/envelope misalignment (see Section \ref{section:I04302}).}
\label{fig:scattered_light}
\end{figure*}

Our model for IRAS 04295+2251 fits both the broadband SED and millimeter visibilities, and naturally reproduces the scattered light image, as seen in Figure \ref{fig:scattered_light}. We have not resolved the disk well, but it appears to have a radius of about 130 AU and a mass of $\sim0.02$ M$_{\odot}$. The best-fit model indicates that the disk has a relatively low inclination ($i\sim60^{\circ}$). The good match to the scattered light image, even though the scattered light image was not used to determine the fit, validates our inferred inclination. The envelope is of comparable mass to the disk, but the visibility profile does not flatten at small $<10$ k$\lambda$ scales, which may indicate that there is large scale envelope material that is resolved out by our observations.

Our best fit model is generally in agreement with what is found by previous studies. \citet{Eisner2005} found that the disk has a radius of 100 AU and that the inclination is low ($i\sim20^{\circ}$). \citet{Eisner2012} found that IRAS 04295+2251 has a compact (30--100 AU) disk with a mass of 0.01 M$_{\circ}$ and a higher inclination, of 45--55$^{\circ}$. \citet{Furlan2008} model the SED and find a very compact ($R_{disk} = 20$ AU) disk with an inclination of 70$^{\circ}$. \citet{Chiang1999} suggested that IRAS 04295 could be an edge on disk, but our modeling indicates that the disk that has a low inclination, and an envelope component is needed to reproduce the observations.

\subsection{IRAS 04302+2247}
\label{section:I04302}

IRAS 04302+2247 is a well-known edge-on disk \citep{Wolf2003,Wolf2008,Eisner2012} nicknamed the ``butterfly star" by \citet{Lucas1997} for it's scattered light morphology, and our modeling results are in agreement with that. Our results suggest that it has a massive disk, with $M_{disk} \sim 0.1$ M$_{\odot}$, and a radius of $\sim240$ AU. Although the envelope is still comparable in mass to most of our targets ($M_{env} \sim 0.02$ M$_{\odot}$), it is several times less massive than the disk, possibly indicating that IRAS 04302+2247 may be in the process of shedding the final layers of it's envelope. Alternatively, it is possible that we are resolving out large scale structure in the envelope, as has been pointed out for several of our other targets.

Although the general morphology of the scattered light image is reproduced by our modeling, the scattered light image prefers a model that is even more edge on \citep[also see][]{Wolf2003} than what we find here ($i \sim 78\pm1$). Interestingly, our best fit model appears to preclude a disk that is precisely edge on, as is suggested by the scattered light morphology. This apparent misalignment of the disk, as traced by millimeter dust emission, and envelope, as traced by scattered light, has been previously noted \citep{Eisner2012}. We speculate that this apparent misalignment may be due to a warped disk, perhaps driven by a massive non-coplanar companion \citep[e.g.][]{Mouillet1997,Dawson2011}, or a misalignment of the disk and envelope, perhaps caused by a perturbation by a passing star sometime in the past \citep[e.g.][]{Quillen2005}.

Our best fit model is in good agreement with the modeling results from \citet{Eisner2012}, which found that the disk has a radius of 250 AU and an inclination of $70-90^{\circ}$. That said, our model suggests that the disk is more massive than their best fit models ($0.005-0.01$ M$_{\odot}$). \citet{Eisner2012}, however, argues that their grid cannot produce a model that fits all of the datasets simultaneously. Our best fit model is also in good agreement with \citet{Wolf2003}, who model the SED, millimeter visibilities and scattered light imaging to find that the disk has a mass of $0.07$ M$_{\odot}$ and a radius of 300 AU. \citet{Gramajo2010} also find a similar disk mass, radius and inclination by fitting the SED and scattered light image, but find a substantially higher envelope mass ($M_{env}\sim0.12$ M$_{\odot}$). Studies that consider just the SED \citep{Kenyon1993a,Whitney1997,Furlan2008} or just the scattered light image \citep{Lucas1997,Stark2006} typically find similar results.

\subsection{IRAS 04365+2535}

IRAS 04365+2535 is one of the few in our sample with a detected Keplerian rotating disk \citep[e.g.][]{Harsono2014,Aso2015}. Our best fit model for it has a disk with a radius of $R_{disk} \sim 140$ AU  and a mass of $0.03$ M$_{\odot}$ embedded in a fairly massive, $\sim0.07$ M$_{\odot}$ envelope of material. The disk appears to be highly inclined $i \sim 55^{\circ}$. The visibility profile is flat from 50 k$\lambda$ onwards, likely indicating the presence of an unresolved disk, but short-ward of this the visibilities rise and trace emission from the envelope. There's no clear evidence of a flattening of the visibilities at short baselines, so it is likely that we have resolved out large scale structure of the envelope. Like IRAS 04016+2610, we need to add a small amount of foreground extinction to fit the near-infrared photometry. This may, however, be because our millimeter visibilities resolve out large scale emission and our model is not correctly capturing the large scale envelope structure.

Our observations are generally in good agreement with results from previous studies. \citet{Chandler1996} suggested that the disk must be inclined by $40-68^{\circ}$ based on observations of IRAS 04365+2535's bipolar outflow, and \citet{Hogerheijde1998} similarly found an inclination of 55$^{\circ}$. Both \citet{Kenyon1993a} and \citet{Whitney1997} modeled the SED and found inclinations of $60^{\circ}$ and $\sim70-90^{\circ}$ respectively. \citet{Whitney1997} also found a disk radius of 50 AU, smaller than we find here. \citet{Gramajo2007} modeled scattered light images of the system and found an inclination of $\sim70^{\circ}$. 

\citet{Harsono2014} observed Keplerian rotation in the IRAS 04365+2535 disk with $^{13}$CO observations and modeled the disk with a radius of 80-100 AU and inclination of $55^{\circ}$. Similarly, \citet{Aso2015} modeled infall and rotation detected towards the prototar in C$^{18}$O emission and found that the disk has an inclination of 65$^{\circ}$ and a radius of 100 AU. These results are both consistent with our own model fits.

Unlike these other studies, though, \citet{Robitaille2007}, \citet{Furlan2008}, and \citet{Eisner2012} all find much lower disk inclinations of $i \sim 18-30^{\circ}$. It is perhaps not surprising that \citet{Robitaille2007} and \citet{Furlan2008} find different inclinations, as they only consider the SED in their modeling. Our results likely differ from \citet{Eisner2012} because their observations did not resolve the disk well. The more recent studies with higher quality millimeter data \citep{Harsono2014,Aso2015}, though, seem to agree with the results presented here.

\section{Discussion}
\label{section:discussion}

\subsection{Class I vs. Class II Disk Masses}

Over the past few decades, there have been numerous studies of nearby star forming regions at millimeter wavelengths with the aim of measuring disk masses for large samples of disks, and this work has been accelerated in recent years by the power of ALMA to quickly survey large numbers of sources \citep[e.g.][]{Beckwith1990,Osterloh1995,Dutrey1996,Andrews2005,Andrews2007,Eisner2008,Mann2010,Andrews2013,Mann2014,Ansdell2016,Eisner2016,Pascucci2016,Barenfeld2016,Ansdell2017}. These surveys have tended to target the population of Class II protostar disks because they are no longer embedded in an envelope, and so estimates of their disk masses are more straightforward. We can compare the Class I disk masses measured here with those of the older Class II disks.

\begin{figure}
\centering
\figurenum{4}
\includegraphics[width=3.2in]{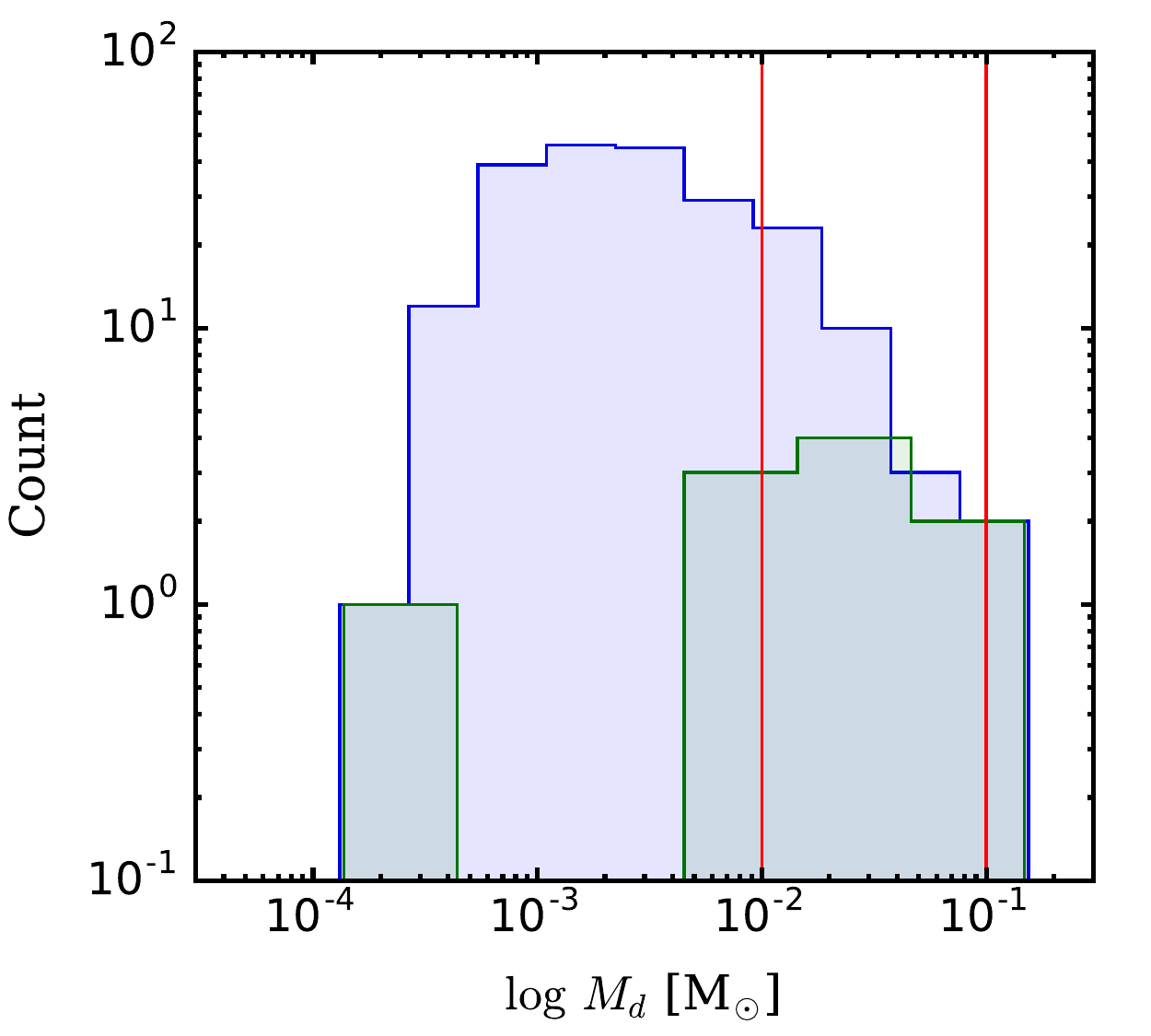}
\caption{Histograms of the disk masses of Class I ({\it green}) sources in our sample and Class II ({\it blue}) sources in Taurus \citep{Andrews2013}. The red lines show the range of lower limits for the Minimum Mass Solar Nebula (e.g. Weidenschilling 1977). We find that our Class I disks, on average, are more massive than the Taurus Class II disks, likely due to dust grain processing hiding matter in larger bodies in the older Class II disks. However, there is still a lack of massive, $>0.1$ M$_{\odot}$ disks, which may be needed to form giant planets.}
\label{fig:histograms}
\end{figure}

We show histograms of disk masses for our sample of Class I disks compared with the sample of Class II disks in Taurus from \citet{Andrews2013} in Figure \ref{fig:histograms}. We show the Taurus Class II disk masses because they are from the same region as our Class I sample, but Class II disks from other regions have similar distributions \citep[see][]{Ansdell2017}. We calculate the disk masses for the Class II sample assuming optically thin dust so that,
\begin{equation}
M_{disk} = \frac{F_{\nu} \, d^2}{\kappa_{\nu} \, B_{\nu}(T)}.
\label{equation:disk_mass}
\end{equation}
We use standard assumptions, of $\kappa_{1.3mm} = 2.3$ cm$^2$ g$^{-1}$ \citep[e.g.][]{Beckwith1990} and $T = 20$ K. We also assume a standard gas-to-dust ratio of 100.

We find that the median Class I disk mass is 0.017 M$_{\odot}$. This is several times higher than the Class II median disk mass, which we find to be 0.0024 $M_{\odot}$ for the Taurus sample. A more detailed study of Class II disk masses finds that the mean Class II disk mass ranges from $0.0015 - 0.0045$ M$_{\odot}$ for a number of nearby star forming regions \citep[see][]{Ansdell2017}. Through the remainder of this section we use the median Class II disk mass that we calculate to compare with our median Class I disk mass. 

To ensure a fair comparison of Class I and II disk masses, however, several precautions must be taken. First, our disk masses are derived using temperature distributions and opacities that are allowed to vary, while our Class II disk masses fix both. To test whether these differences are affecting our results, we have computed a disk-only 1.3 mm flux for each of our sources by raytracing their models with the envelope removed. Disk masses are then computed from those fluxes using Equation \ref{equation:disk_mass}. We find that the median Class I disk mass when computed this way is 0.015 M$_{\odot}$, only slightly different from our original calculation. The slightly lower disk mass makes sense, as several of our sources have disks that are optically thick in the inner regions.

It has also been shown that Class II disk masses are correlated with stellar mass \citep{Andrews2013,Barenfeld2016,Pascucci2016,Ansdell2016}. As such, when comparing disk mass distributions, we must take care to ensure that our samples have similar stellar host properties; otherwise our disk mass measurements may be biased. We have used spectral types measured for 7 of our targets \citep{Doppmann2005,Connelley2010} and a direct mass measurement of an eighth to estimate stellar host masses for our sample. We compare these masses with the Class II stellar masses using  two-sample tests and find that $p = 0.1 - 0.19$, indicating that we cannot distinguish between the stellar populations. As such, we expect that it is fair to directly compare disk masses. When we compare the two disk mass distributions below, however, we do account for the underlying stellar mass distribution in one test.

If we assume that the disk masses for Class I and II protostars are distributed normally in log-space, which may not be a good assumption but enables a simple comparison of the two samples, then a two-sided t-test finds a probability of $p = 0.018$ that they are drawn from distributions with the same mean value. If, instead, we split each sample up into two categories, disks above and below the median Class II disk mass, then a Fisher Exact test, which is distribution independent, finds a probability of $p = 0.019$ that Class I and Class II disks are drawn from the same distribution. Thus the disk mass distributions among Class I and II sources appear different, with a significance of $>2$ sigma. Finally, we follow the procedure outlined in \citet{Andrews2013} to compare the distributions: First we randomly draw 10 disks from the Class II sample. Eight disks are drawn to have the same host properties as the eight Class I's with known spectral types, and two are drawn from the entire distribution to represent the two Class I's with unknown spectral types. Next, we compute two sample tests such as the Kolmogorov-Smirnov test or the log-rank test to compute the null-hypothesis probability. Finally, we repeat this process a large number of times while recording the null-hypothesis probability each time. We find that the two samples are different at the $>2\sigma$ level in 81\% of the trials and different at the $>3\sigma$ level in 45\% of the trials.

Our sample is missing 2 of the 12 companionless bona-fide Class I protostars in Taurus, and those two are among the faintest of our targets when when observed with a single dish telescope \citep[e.g.][]{Motte2001}. If their faintness also corresponds to a low disk mass, it is possible that we may be artificially boosting the median disk mass of Class I sources by biasing our sample towards higher mass disks. If we assume that both sources are similar in mass to IRAS 04181+2654A, which is at the low end of our disk mass distribution, however, we still calculate a median disk mass of 0.011 M$_{\odot}$. It is important to note, however, that a lower single-dish millimeter flux may not indicate a low-mass disk. IRAS 04108+2803B has a comparable single dish flux to both of these objects \citep{Motte2001} and yet we find that its disk mass is near the median for Class I disks.

The higher average mass of Class I disks compared with Class II disks is an indication that substantial dust processing and grain growth occurs between the Class I and II stages. If dusty disk material has grown into rock, planetesimal, and planet sizes by the Class II stage, then much of this matter would be hidden from millimeter surveys, which are primarily sensitive to millimeter sized dust. For a dust grain size distribution $n \propto a^{-p}$, the opacity varies with maximum grain size like $\kappa \propto a_{max}^{p-4}$ \citep{Draine2006}. In order to explain the factor of $\sim10$ decrease in dust mass from the Class I to the Class II stage as a change in opacity, a decrease in millimeter opacity by a factor of $\sim10$ is needed. As our models assume $p = 3.5$, an increase in the maximum dust grain size by a factor of $\sim75$, say from 1 mm to 10 cm, could explain this drop in flux. This is borne out by a number of studies that have found cavities, gaps, spiral arms and other asymmetries in Class II disks that may indicate the presence of planets \citep{Isella2010,Andrews2011,vanderMarel2013,Casassus2013,Andrews2016,Perez2016,Isella2016,Loomis2017,Fedele2017}, although planets have so far only been found in a few disks \citep[e.g.][]{Sallum2015}.

\subsection{Implications for Giant Planet Formation}

Recent disk mass surveys of Class II protostars have raised concerns about whether their disks contain enough mass to form giant planets \citep[e.g.][]{Williams2014,Eisner2016,Ansdell2016}. An accounting of the material in our own Solar System, which is dominated by the mass of Jupiter, suggests that disk masses of $\gtrsim0.01-0.1$ M$_{\odot}$ are needed to form a planetary system like our own \citep[e.g.][]{Weidenschilling1977,Hayashi1981,Desch2007}. The masses inferred from sub-millimeter observations of Class II disks are, on average, below this Minimum Mass Solar Nebula. It has also been found recently that gas-to-dust ratios in Class II disks may be well below the canonical value of 100 \citep{Williams2014,Eisner2016,Ansdell2016}. If true, this would create further discrepancies with the MMSN, although it may simply be that CO is depleted in Class II disks \citep[e.g.][]{Miotello2017}.

Whether Class II disks have enough mass to form giant planets may, however, be irrelevant, as evidence is mounting that planets are already present in Class II disks (see above). As such, the Class I disks, which are younger \citep[e.g.][]{Evans2009,Dunham2015} and have had less time for dust processing and planet formation to occur, should better represent the initial mass budget of disks for forming planets. And although the Class I disk sample appears to be more massive, on average, than the Class II sample, it remains unclear from our results whether Class I disks are massive enough to form giant planets. 

With a median disk mass of 0.016 M$_{\odot}$, Class I disks do have enough mass, on average, to form giant planets if the minimum amount of matter needed is 0.01 M$_{\odot}$. However this median disk mass is still well below the MMSN estimates of 0.06 M$_{\odot}$ \citep{Desch2007} and the high end of 0.1 M$_{\odot}$ \citep{Weidenschilling1977}. A t-test shows with $>2\sigma$ confidence ($p = 0.03$) that the mean Class I disk mass is below 0.06 M$_{\odot}$ and with $\sim3\sigma$ confidence ($p = 0.007$) that the mean is below 0.1 M$_{\odot}$. There are two sources (i.e. 20\% of the sample; IRAS 04158+2805 and IRAS 04302+2247) that have $M_{disk} \gtrsim 0.06 - 0.1$ M$_{\odot}$, comparable to the $\sim20\%$ of stars with giant planets \citep{Cumming2008}, but this is clearly not statistically significant.

If the upper end of the MMSN estimates do represent better estimates of the initial amount of matter needed to form giant planets, this may be an indication that planet formation has already begun during the Class I stage. In fact, recent observations with ALMA provide evidence that this is the case. The HL Tau system, which is now known to have a series of narrow gaps in it's disk \citep[e.g.][]{Brogan2015}, is thought to be somewhere between the Class I and II stages and is likely $\sim1$ Myr old. If the gaps are carved by planets \citep{Dong2015}, it would be an indication that planet formation must begin early enough to form Saturn-mass planets \citep[e.g.][]{Dong2015,Kanagawa2015} within the first $\sim$ Myr. Perhaps even more interesting, several Class I protoplanetary disks have recently been found to also exhibit similar features. This includes WL 17, which has a 12 AU-wide hole in the center of its disk \citep{Sheehan2017a}, and GY 91, which has three narrow dark lanes and is very similar to the HL Tau disk (Sheehan \& Eisner, in prep.). Although these features could very well be produced by something other than planets, many of the likely causes are still indications that the planet formation process has begun. If planet formation occurs during the Class I stage then we would expect that disk masses are even higher at younger ages, perhaps during the Class 0 stage, before dust processing has had time to progress significantly.


\section{Conclusions}
\label{section:conclusion}

We have presented an updated method for fitting disk+envelope radiative transfer models to a multi-wavelength dataset \citep[e.g.][]{Eisner2005,Eisner2012,Sheehan2014} that uses Markov Chain Monte Carlo fitting. Although these models are computationally intensive to run, the fitting can be done in a reasonable amount of time when run in parallel on systems with a large number of cpus. 



We have used this modeling infrastructure to fit disk+envelope models to a sample of 10 Class I protostars in the Taurus Molecular Cloud. These sources were chosen because they are widely accepted to be Class I objects and also because none have been found to have close companions. We find good fits to the combined broadband SED and CARMA 1.3 mm visibilities dataset for each source. The resulting best fit models are even good matches to HST scattered light images, when such images are available, despite the fit not including these data.

From our best fit models we are able to determine the disk masses for this sample of Class I sources. We find that the median Class I disk mass is 0.017 M$_{\odot}$, which is higher than the median Class II disk mass by a factor of a few, although it remains unclear whether Class I disks have enough mass in millimeter-sized dust grains, on average, to form giant planets. Larger samples of Class I disks are needed to better nail down the Class I disk mass distribution With ALMA now online, a much larger sample of Class I disks can be observed with higher spatial resolution and better sensitivity far more efficiently, so it is only a matter of time before these questions are answered.

\acknowledgements

We would like to thank Nick Ballering for many useful conversations regarding our MCMC fitting procedure, and Dan Marrone for helping to expedite the acquisition of our CARMA dataset. We'd also like to thank our anonymous referee for comments which helped to improve the paper.
This material is based upon work supported by the National Science Foundation Graduate Research Fellowship under Grant No. 2012115762.
This work was supported by NSF AAG grant 1311910.
The results reported herein benefitted from collaborations and/or information exchange within NASA's Nexus for Exoplanet System Science (NExSS) research coordination network sponsored by NASA's Science Mission Directorate.
Support for CARMA construction was derived from the Gordon and Betty Moore Foundation, the Kenneth T. and Eileen L. Norris Foundation, the James S. McDonnell Foundation, the Associates of the California Institute of Technology, the University of Chicago, the states of California, Illinois, and Maryland, and the National Science Foundation. CARMA development and operations were supported by the National Science Foundation under a cooperative agreement, and by the CARMA partner universities.

\bibliography{ms.bib}

\begin{figure*}
\centering
\figurenum{5b}
\includegraphics[width=7in]{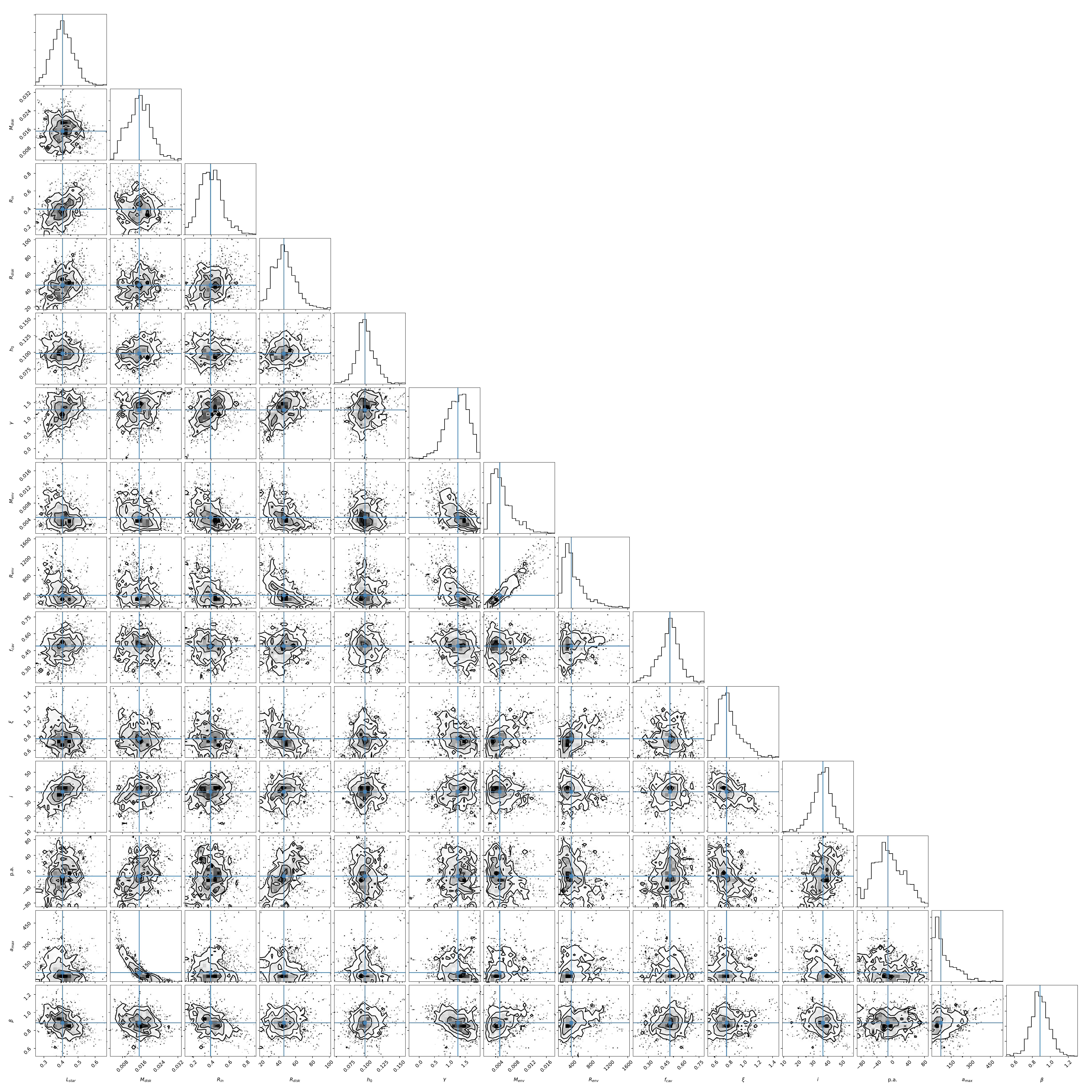}
\caption{Projected posterior pdf for every combination of parameter pairs for IRAS 04108+2803B.}
\end{figure*}

\begin{figure*}
\centering
\figurenum{5c}
\includegraphics[width=7in]{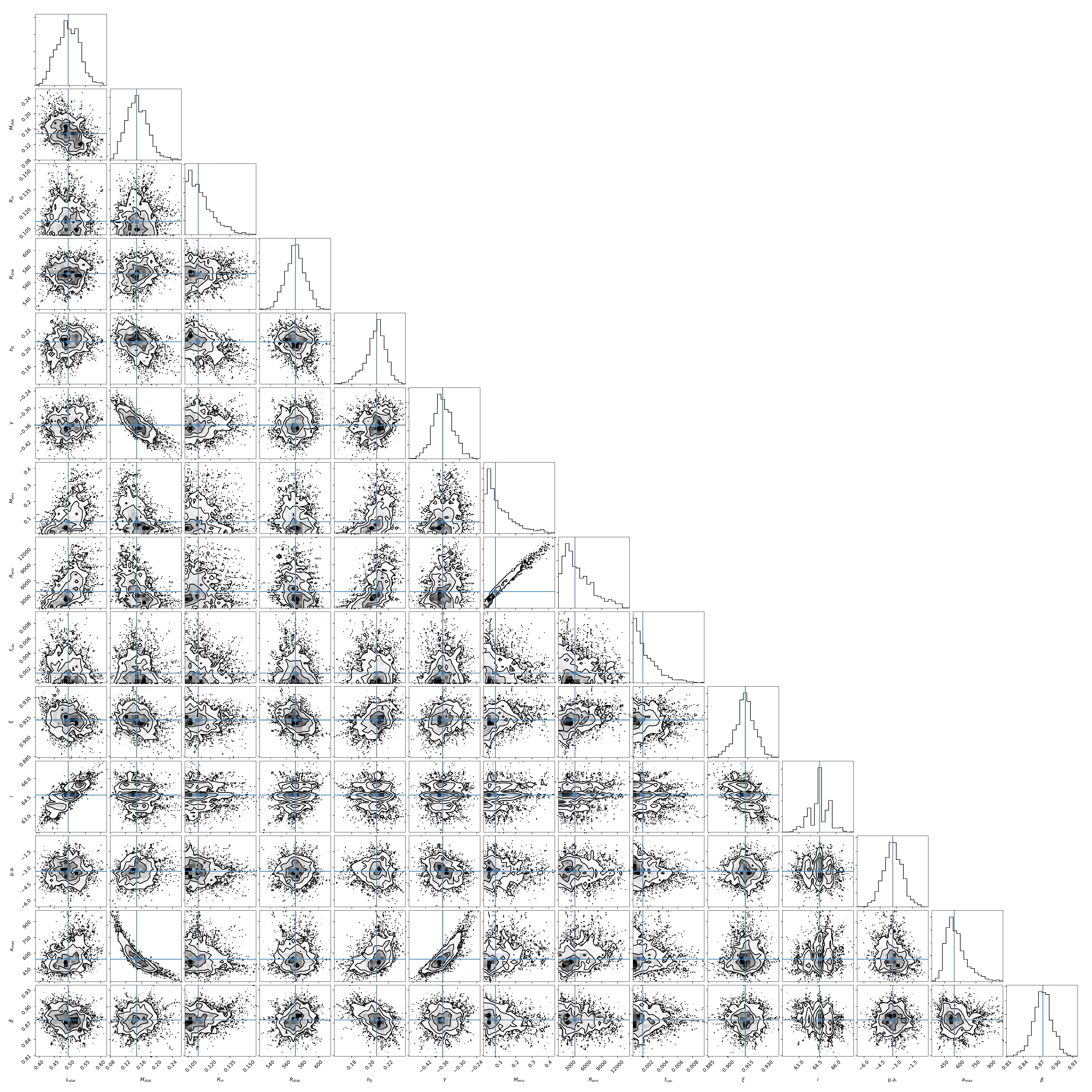}
\caption{Projected posterior pdf for every combination of parameter pairs for IRAS 04158+2805.}
\end{figure*}

\begin{figure*}
\centering
\figurenum{5d}
\includegraphics[width=7in]{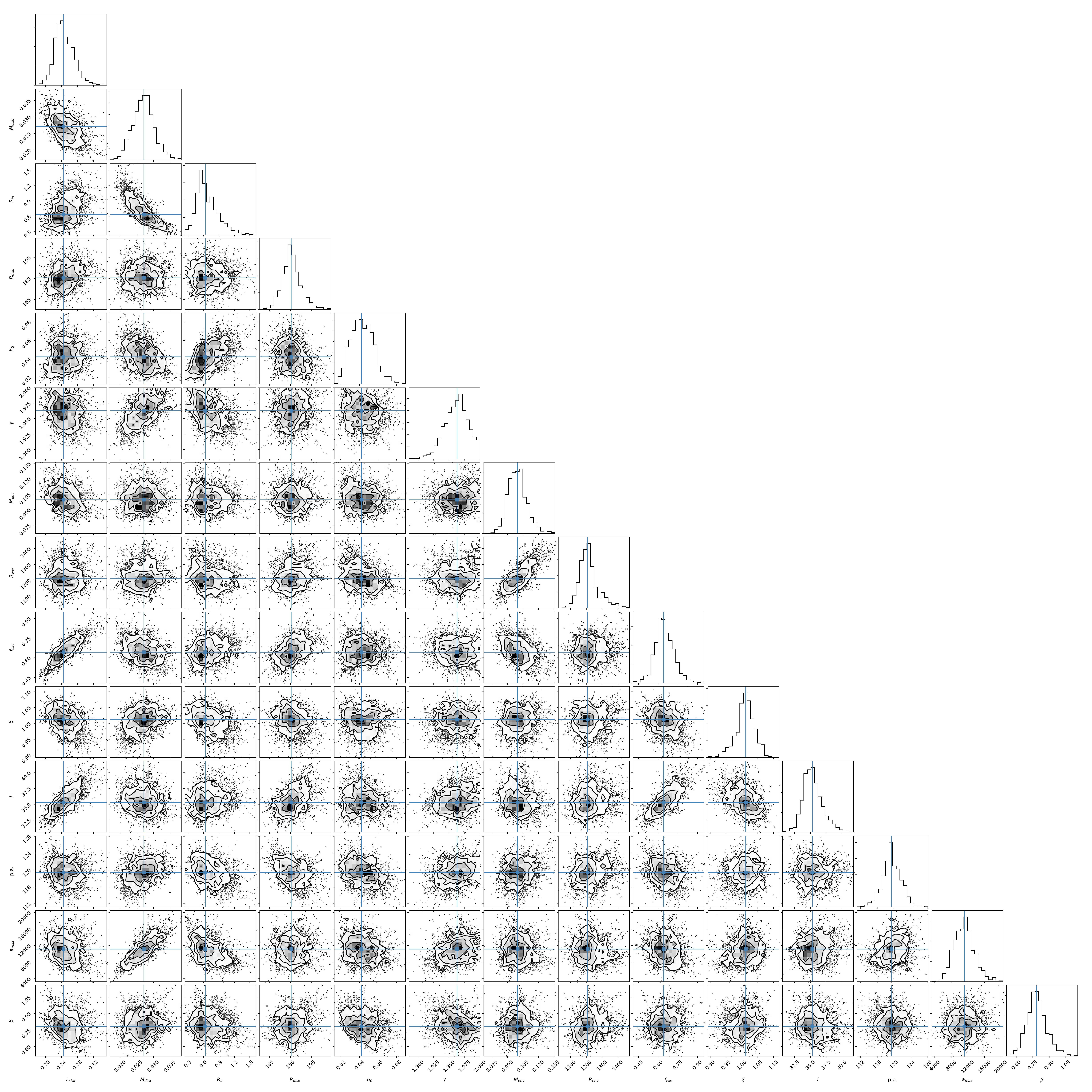}
\caption{Projected posterior pdf for every combination of parameter pairs for IRAS 04166+2706.}
\end{figure*}

\begin{figure*}
\centering
\figurenum{5e}
\includegraphics[width=7in]{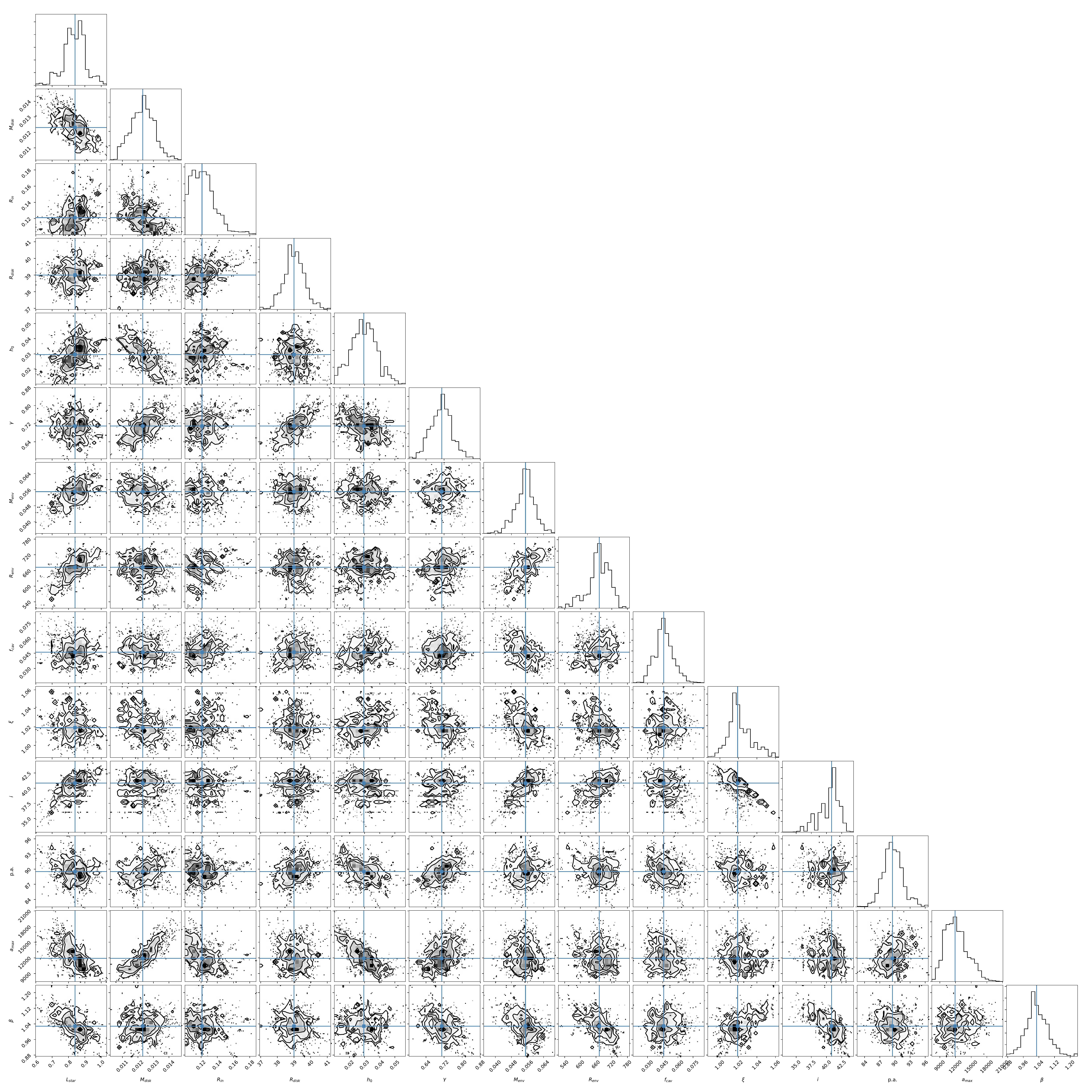}
\caption{Projected posterior pdf for every combination of parameter pairs for IRAS 04169+2701.}
\end{figure*}

\begin{figure*}
\centering
\figurenum{5f}
\includegraphics[width=7in]{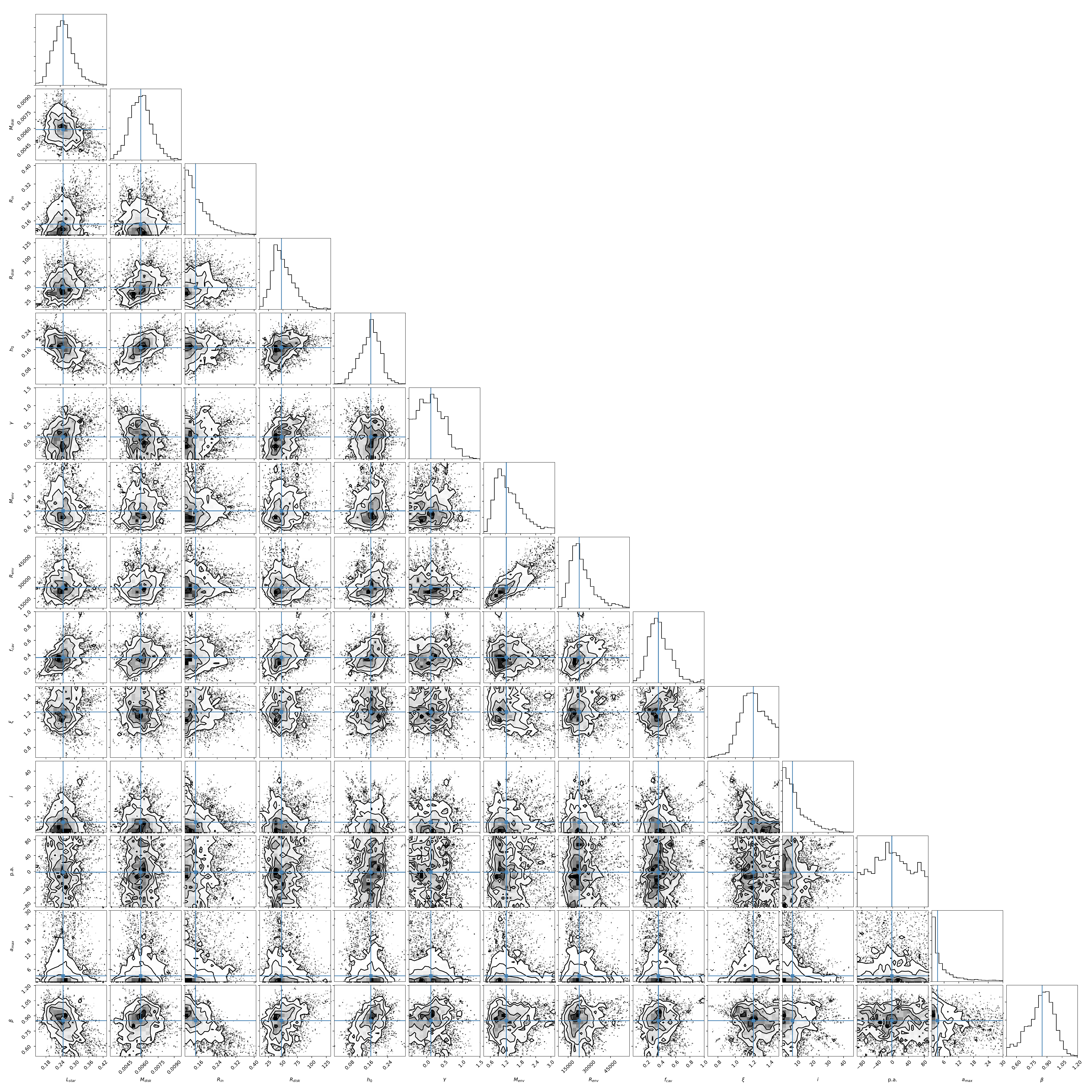}
\caption{Projected posterior pdf for every combination of parameter pairs for IRAS 04181+2654A.}
\end{figure*}

\begin{figure*}
\centering
\figurenum{5g}
\includegraphics[width=7in]{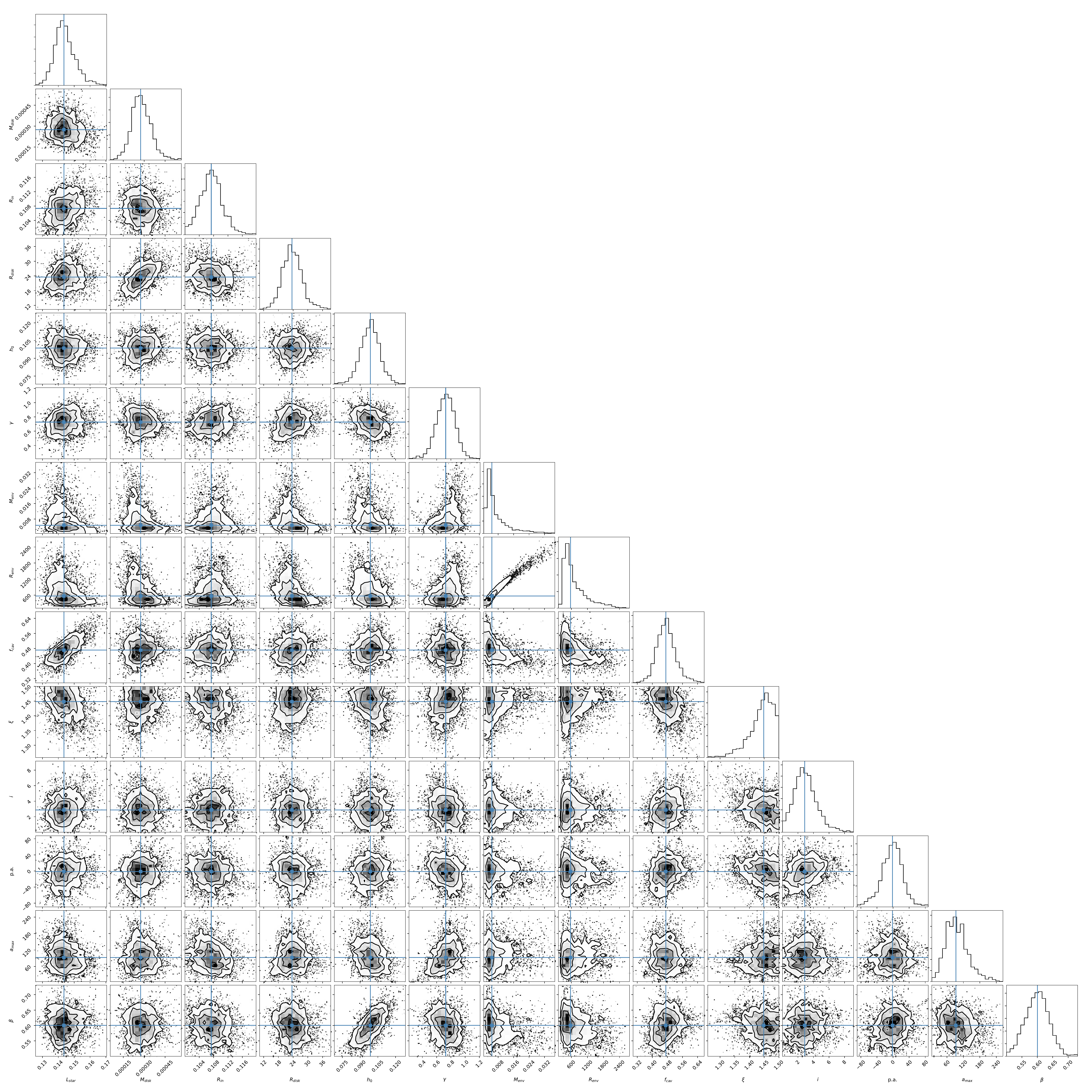}
\caption{Projected posterior pdf for every combination of parameter pairs for IRAS 04181+2654B.}
\end{figure*}

\begin{figure*}
\centering
\figurenum{5h}
\includegraphics[width=7in]{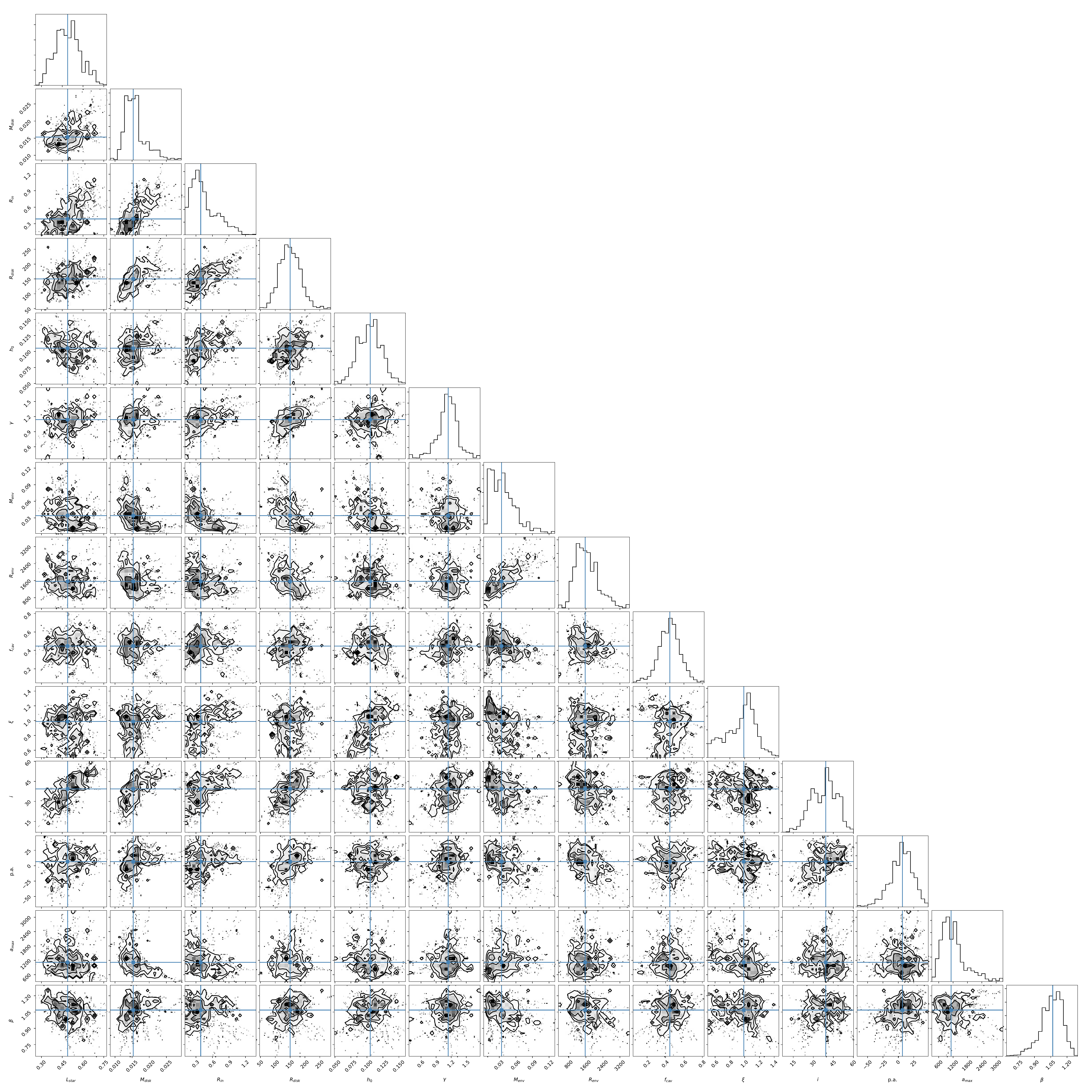}
\caption{Projected posterior pdf for every combination of parameter pairs for IRAS 04295+2251.}
\end{figure*}

\begin{figure*}
\centering
\figurenum{5i}
\includegraphics[width=7in]{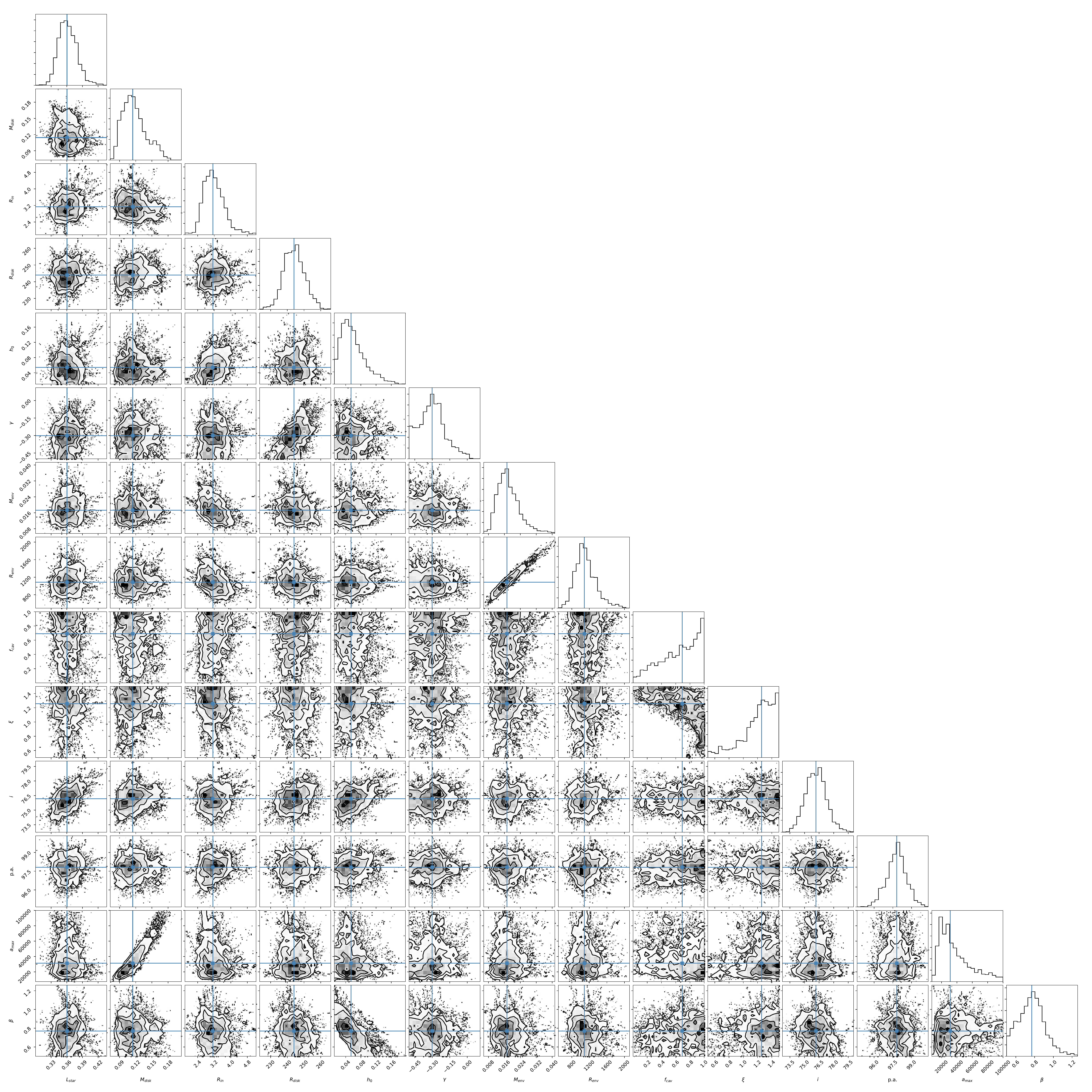}
\caption{Projected posterior pdf for every combination of parameter pairs for IRAS 04302+2247.}
\end{figure*}

\end{document}